\documentclass[%
 reprint,
 amsmath,amssymb,
 aps,
]{revtex4-2}
\usepackage{graphicx}
\usepackage{dcolumn}
\usepackage{bm}
\usepackage{color}

\def\mnras{Mon. Not. R. Astron. Soc. }

\def\apj{Astrophys. J.}
\def\apjl{Astrophys. J. Lett.}
\def\prd{Phys. Rev. D}
\def\prl{Phys. Rev. Lett.}

\def\aap{Astron. Astrophys.}

\def\araa{Annu. Rev. Astron. Astrophys.}

\def\aj{Astron. J.}

\begin{document}


\title{
Detecting Accelerating Eccentric Binaries in the LISA Band 
}

\author{Zeyuan Xuan}
\email{Corresponding author: zeyuan.xuan@physics.ucla.edu}
\affiliation{ Department of Physics and Astronomy, UCLA, Los Angeles, CA 90095}
\affiliation{Mani L. Bhaumik Institute for Theoretical Physics, Department of Physics and Astronomy, UCLA, Los Angeles, CA 90095, USA}

\author{Smadar Naoz}
\email{snaoz@astro.ucla.edu}
\affiliation{ Department of Physics and Astronomy, UCLA, Los Angeles, CA 90095}
\affiliation{Mani L. Bhaumik Institute for Theoretical Physics, Department of Physics and Astronomy, UCLA, Los Angeles, CA 90095, USA}

\author{Xian Chen}
\email{xian.chen@pku.edu.cn}
\affiliation{Astronomy Department, School of Physics, Peking University, 100871 Beijing, China}
\affiliation{Kavli Institute for Astronomy and Astrophysics at Peking University, 100871 Beijing, China}

\date{\today}

\begin{abstract}
Many gravitational wave (GW) sources in the LISA band are expected to have non-negligible eccentricity. Furthermore, many of them can undergo acceleration because they reside in the presence of a tertiary. Here we develop analytical and numerical methods to quantify how the compact binary's eccentricity enhances the detection of its peculiar acceleration. We show that the general relativistic precession pattern can disentangle the binary's acceleration-induced frequency shift from the chirp-mass-induced frequency shift in GW template fitting, thus relaxing the signal-to-noise ratio requirement for distinguishing the acceleration by a factor of $10\sim100$. Moreover, by adopting the GW templates of the accelerating eccentric compact binaries, we can enhance the acceleration measurement accuracy by a factor of $\sim100$, compared to the zero-eccentricity case, and detect the source's acceleration even if it does not change during the observational time. For example, a stellar-mass binary black hole (BBH) with moderate eccentricity in the LISA band yields an error of the acceleration measurement $\sim10^{-7}m\cdot s^{-2}$ for $\rm{SNR}=20$ and observational time of $4$~yrs. In this example, we can measure the BBHs' peculiar acceleration even when it is $\sim1\rm pc$ away from a $4\times 10^{6}\rm M_{\odot}$ SMBH. Our results highlight the importance of eccentricity to the LISA-band sources and show the necessity of developing GW templates for accelerating eccentric compact binaries.
\end{abstract}

\maketitle

\section{Introduction}
\label{section:intro}
The detection of gravitational wave (GW) by LIGO
and Virgo \citep{2021ApJ...913L...7A,2021arXiv211103634T} has greatly enhanced our understanding of the properties of compact objects in the Universe. In the future, the Laser Interferometer Space Antenna (LISA) \citep{2017arXiv170200786A} would broaden the scope of GW astronomy by detecting GW signals in a lower frequency band ($10^{-4}-10^{-1}~\rm Hz$), where GW signal in the LISA band may last for years \citep{amaro+22}, potentially having signatures of compact binary's long-term evolution long before they merge. Therefore, LISA detection can shed light on compact object binary's formation channel and surrounding environment.

Among all the parameters that can be measured using LISA, the GW source's acceleration is of great importance, because it carries the signature of the binary's environment. Several studies suggested that accelerating compact object binaries may significantly contribute to the population observed by LISA \citep{Pribulla+06,Tokovinin+08,Raghavan+10,Sana+11,Sana+12,Moe+17,Naoz+13,Naoz+14,Hamers+16,Hamers+18,Hamers+20,Rose+19,Hamers+13, Toonen+16, Toonen+18,Toonen+17, Perpiny+19,lagos20,O'Leary+09,Antonini+12,Antonini+14,Stephan+16,Stephan+19,Hoang+18,Wang+21}.
For example, observational endeavors shows that stars often reside in binaries, 
and high mass stars reside in higher multiples \citep{Pribulla+06,Tokovinin+08,Raghavan+10,Sana+11,Sana+12,Moe+17}. According to long term stability arguments, the spatial separation of these multiples is likely to evolve into one tight inner binary with tertiary in wide outer orbit(s) \citep{Naoz+13,Naoz+14,Hamers+16,Hamers+18,Hamers+20,Rose+19}. Stellar evolution of such systems may result in compact object binaries that host at least a tertiary companion \citep{Hamers+13, Toonen+16, Toonen+18}.
  The observation of white dwarfs (WDs) also directly suggests that many WDs binaries reside in a triple configuration \citep{Toonen+17, Perpiny+19,lagos20}. These compact object binaries can undergo non-negligible acceleration in the triple system, thus leaving an imprint of acceleration on the GW signal.

Additionally, many binaries and compact object binaries are supposed to reside in the center of galaxies \citep{O'Leary+09,Antonini+12,Antonini+14,Stephan+16,Stephan+19,Naoz+18,Hoang+18,Wang+21}. Already a few stellar binaries were detected in the inner $\sim 0.1$~pc of our galactic center \citep{Pfuhl+14}. 
Moreover, the number of compact binaries visible in the LISA band within the inner parsec of our galactic center is estimated to be $14 - 150$ WD-WD, $0 - 2$ neutron star (NS) - black hole (BH), $0.2 - 4$ NS-NS, $0.3 - 20$ BH-BH, and the results are of the same order for other galaxies \citep{Wang+21}. These systems mentioned above can undergo non-negligible acceleration caused by the gravitational potential of the supermassive black hole (SMBH) in the galactic center.

Measuring the acceleration will give us direct evidence about the GW source's dynamic environment. For example, a binary black hole (BBH) orbiting the SMBH on a close configuration will undergo a time-dependent Doppler shift induced by acceleration, which is potentially detectable \citep{Inayoshi+17,Meiron+17,Robson+18,Randall+19Dop,Wong+19,Gupta+20,Tamanini+20}. Additionally, an acceleration signature on GW signal may help detect double WDs (DWDs) accompanied by a star or a planet \citep{Seto+08,Bonvin+17,Robson+18,Steffen+18,Tamanini+19,Danielski+19}. Further, eccentricity oscillations due to the Eccentric Kozai Lidov mechanism \citep[e.g.,][]{kozai62,Lidov1962,Naoz16}, can induce a time-changing characteristic strain profile in the LISA band \citep{Hoang+19,Randall+19,Deme+20,Emami+20}, thus imprinting a signature on a binary in the presence of a tertiary.
However, the effect of acceleration on the GW signal is not always distinguishable and measurable. Sometimes it may degenerate with the measurement of other parameters and yield a misleading interpretation of the physics behind the GW sources \citep{Robson+18,Tamanini+20,chen20envi,Xuan+21}.

Consider a circular compact object binary as an example. When its evolution is dominated by GW radiation and the observer is in the rest frame of the source, the intrinsic GW frequency $f_{e}$ will increase with time. The frequency shift rate $\dot{f}_{e}$ is proportional to the 
``chirp mass'' as: $\dot{f}_{e}\sim {\cal M}_{c}^{5/3}$, where ${\cal
M}_{c}\equiv (m_1m_2)^{3/5}/(m_1+m_2)^{1/5}$, and $m_1$
and $m_2$ are masses of the two compact objects \citep{Cutler+94}. 
But if the GW source is accelerating, the peculiar acceleration also leads to an extra frequency shift rate $\dot{f}_{\rm acc}$ by changing the peculiar velocity of the GW source and inducing a time-dependent Doppler shift. 
Therefore, the detected GW signal has two components that contribute to the observed frequency shift rate, one from the GW emission that shrinks and circularizes the orbit, and the other from the acceleration of the system. 
In other words, there is a degeneracy between the chirp mass and peculiar acceleration since they both contribute to the frequency shift rate in the GW signal for the leading order. 

Such kind of degeneracy makes it harder to measure peculiar acceleration. Some studies suggested that 
GW source with measurable acceleration may be limited to the cases when the compact binary is very close to the tertiary and the period of the outer orbit is short enough \citep{Robson+18,Tamanini+20}. 
Further, for DWDs in the Milky Way, there is a significant parameter space where acceleration is large enough to cause a non-negligible $\dot{f}_{\rm acc}$ yet still not distinguishable (degenerate with other parameters) in GW template fitting \citep{Xuan+21}.

However, there can be a different story when the eccentricity of the GW source is detected. In fact, several studies suggest that many eccentric compact binaries are in the LISA band \citep{O'Leary+09,Samsing+18,Hoang+19,Fragione+19,breivik20,Naoz+20,Wang+21,Zhang+21}.
And in some dynamical channels, the existence of a tertiary is supposed to directly produce LISA-band sources by exciting the eccentricity of inner orbit and accelerating the merger \citep{Thompson+11,Antognini+14,Hoang+18,Stephan+19,Martinez+20,Hoang+20,Naoz+20,Stephan+19,Wang+21}. Thus, we expect binaries in an accelerating environment to be eccentric.

If the binary has non-zero eccentricity, the general relativistic (GR) precession will induce a triplet waveform, which in turn changes the frequency-peak position of each harmonic. This signature can be used to extract the binary's total mass independently of the frequency shift rate \citep{Seto+01,Mikoczi+12}, making it possible to break the degeneracy between peculiar acceleration and the chirp-mass-induced frequency shift.

Here we propose a strategy aimed to break this degeneracy by considering the GR precession signature of the accelerating eccentric binary. In Section \ref{sec:analytical}, we present analytical methods yielding an overall understanding of how eccentricity can disentangle the acceleration feature from the chirp-mass-induced frequency shift. In Section \ref{Numerical}, we introduce the numerical tools used to simulate LISA event waveforms (\S \ref{NumericalA}), analyze the waveform (\S \ref{NumericalB}), and estimating the error of parameter measurements (\S \ref{NumericalC}).   Section \ref{examples} shows the application of numerical methods. In particular, in this section, we map the parameter space where the GW signal from accelerating sources can be distinguished from non-accelerating GW templates and quantify the effect of eccentricity on the accuracy of peculiar acceleration measurement. Finally, in Section \ref{conclusion}, we offer our discussion and conclude that eccentric binaries have a clear signature on the GW form when it is undergoing peculiar acceleration. Specifically, the acceleration measurement enhanced by eccentricity can shed light on the GW source's environment. 

\section{Analytical Consideration}\label{sec:analytical}
We begin with establishing the effects of acceleration on the GW signal (\S  \ref{sec:AccGen}). This effect causes a degeneracy between the peculiar acceleration and the chirp-mass-induced frequency shift. We suggest that eccentricity can be used to disentangle this degeneracy. 
The role of eccentricity in detecting compact binaries' peculiar acceleration can be divided into two parts. 
\begin{enumerate}
    \item {\it Ignoring the possible contribution of acceleration in the signal.} In this case, consider using GW templates, in the data analysis, that do not include the binary's peculiar acceleration.  Below (\S \ref{sec:disentangle}), we show that eccentricity can help distinguish an accelerating eccentric compact binary from non-accelerating ones.
    \item {\it Including the possible contribution of acceleration in the signal.} In this case, the data analysis uses the GW templates with accelerating features, making it possible to measure the binary's acceleration.  Below  (\S \ref{sec:MeasureAcc}), we show how the measurement of acceleration depends on the eccentricity of the binary.  Later, using numerical analysis, we demonstrate that the eccentricity can increase the acceleration measurement accuracy (\S \ref{egB}).
\end{enumerate}

\subsection{Mass-Acceleration Degeneracy in GW Data Analysis }\label{sec:AccGen}
Consider a compact object binary with a semi-major axis $a$, eccentricity $e$, and the two components' masses $m_1$ and $m_2$. Its energy is dissipated by the GW radiation, which results in the decrease of the orbital period $P_{b}$.
The average changing rate of the semi-major axis, $a$ is \citep{Peters64}:
\begin{equation}
\left\langle\frac{d a}{d t}\right\rangle=-\frac{64}{5} \frac{G^{3} m_{1} m_{2}\left(m_{1}+m_{2}\right)}{c^{5} a^{3}\left(1-e^{2}\right)^{7 / 2}}\left(1+\frac{73}{24} e^{2}+\frac{37}{96} e^{4}\right) \ ,
\end{equation}
where $G$ is the gravitational constant and $c$ is the speed of light. 

When the binary has a non-zero eccentricity, the GW signal is made up of multiple harmonics. The frequency of GW's n-th harmonic is related to the binary's orbital period as $f_{e}^{j=n}\approx n/P_{b}$, where the subscript $``e"$ denotes quantities in the rest frame of the source.
Note that since the orbital period is shrinking, the frequency of the GW signal will increase with time. This phenomenon is known as the ``chirp signal'' of a GW source.  

GW data analysis studies often focus on the second harmonic of the GW signal, which is the dominant signal for circular and low eccentricities orbits. Thus, here we will take $f_{e}^{j=2}$ as an example to demonstrate the commonly used method for estimating the mass of the GW source. It should be noted that GW data analysis methodology often proposes to fit the numerical template for the parameter estimation of GW sources \citep{Finn92,Cutler+94}, while the analysis presented here aims to explain different factors' contribution to the result of the numerical fitting and give an analytical estimation of the acceleration measurement accuracy in different cases.

The time derivative of the 2nd harmonic is uniquely determined by the frequency of the GW signal, eccentricity, and the mass of the GW source for the leading order:
\begin{equation}
	\dot{f}_{e}^{j=2}=
    \frac{96 \pi^{8/3}}{5}\left(\mathcal{M}_{c}\frac{G}{c^{3}}\right)^{5/3}\left(f_{e}^{j=2}\right)^{11/3}F(e) \ ,
    \label{eq:chirp mass}
\end{equation}
We emphasize that $\mathcal{M}_{c}$ is the intrinsic chirp mass of the binary as a GW source, which only depends on the mass of the compact binary's components; The observed chirp mass, $\mathcal{M}_{o}$, can be different from $\mathcal{M}_{c}$ because the observed GW signal can be distorted. $F(e)$ is a function of the compact binary's eccentricity (the enhancement function)\citep{peters63}:
\begin{equation}\label{eq:Fe}
F(e)\equiv\frac{1+\frac{73}{24}e^{2}+\frac{37}{96}e^{4}}{(1-e^{2})^{7/2}} \ .
\end{equation}
Thus,  we can estimate the intrinsic chirp mass ${\cal M}_{c}$ of compact binaries based on two observables, the
frequency $f_e$ and its time derivative $\dot{f}_e$. 
For the circular binary case, the details of this method can be found in \citet{Cutler+94}. On the other hand, for an eccentric binary, the amplitude profile of different harmonics is related to the eccentricity \citep{Moreno+95,Gopakumar+02,barack04}. It can enable us to measure the eccentricity in template fitting \citep{Cornish+03}. In this way, the results of \citep{Cutler+94} can be generalized by plugging in $f_{e},\dot{f}_{e}$, and $e$ into Eq.~(\ref{eq:chirp mass}) and finding the intrinsic chirp mass of the source :
\begin{equation}
	{\mathcal{M}_{c}}
	={ \left(
\frac{5 c^5}{96 \pi^{8/3} G^{5/3} } \right) }^{3/5} (f^{j=2}_e)^{-11/5} (\dot{f}^{j=2}_e)^{3/5} F(e)^{-3/5} \ .
\label{eq:chirpmass-f}
\end{equation}
When the GW source is moving relative to the observer with a velocity $v$, the frequency of GW's each harmonic will be shifted because of the Doppler effect. In the observer frame, the frequency, $f_{o}$, is then \citep{Meiron+17,Inayoshi+17,Robson+18,chen+19redshift}:
\begin{equation}
f_{o}^{j=n} = f_{e}^{j=n} \frac{\sqrt{1-\beta^{2}}}{1+\beta \cos{\theta}} = f_{e}^{j=n}(1+z_{\rm dop})^{-1} \, ,
\label{eq:doppler}
\end{equation}
where $\beta = v/c$, $\theta$ is the angle between the velocity vector and the line of sight, and $z_{\rm dop}$ is the Doppler coefficient. 

We are interested in the case when the GW source is accelerating, which means that the peculiar velocity and Doppler shift factor are changing with time. Thus, from Eq.~(\ref{eq:doppler}) it is straightforward to find the time derivative of the observed frequency \citep{chen19}, i.e., 
\begin{equation}
\frac{d}{dt}{f}_{o}^{j=n} = \frac{\dot{f}_{e}^{j=n}}{(1+z_{dop})^{2}}+f_{e}^{j=n}\frac{d}{dt}\left(\frac{\sqrt{1-\beta^{2}}}{1+\beta \cos{\theta}}\right) \ ,
\label{eq:doppler2}
\end{equation}
where the power $2$ factor on the Doppler coefficient comes due to the transformation from the observed frames to the source frame.  
The last term of Eq.~(\ref{eq:doppler2}), represents the line of sight acceleration $a_{//}/c$. In other words :
\begin{equation}\label{Eq:acc}
 \frac{a_{//}}{c} = \frac{d}{dt} \left(\frac{\sqrt{1-\beta^{2}}}{1+\beta \cos{\theta}}\right) \ .
\end{equation}
Equivalently, from Eq.~(\ref{eq:doppler2}), this acceleration relates to the difference between the time derivative of frequency in the observer's frame and the rest frame of the GW source. Such difference is directly caused by the change of peculiar velocity, thus can be used to constrain the line of sight acceleration $a_{//}$, i.e.,
\begin{equation}
    \frac{a_{//}}{c}= \frac{\dot{f}_{o}-\dot{f}_{e}(1+z_{\rm dop})^{-2}}{f_{e}} \ .
    \label{eq:acc1}
\end{equation}

However, we cannot directly measure the frequency ($f_{e}$) and its time derivative ($\dot{f}_{e}$ ) in the source frame. Instead, the observed quantities are $f_{o}$ and $\dot{f}_{o}$ in the observer frame. As mentioned above, even when $a_{//}=0$, constant peculiar velocity and cosmological redshift can cause a significant difference between $f_{o}$ and $f_{e}$ and distorts the parameters of the GW source.  This phenomenon is known as ``mass-redshift degeneracy," \citep[e.g.,][]{Sathyaprakash+09,broadhurst18,smith18,chen+19redshift}.
Because of the scale-free property of gravity, the effect of mass-redshift degeneracy can be exactly canceled out by making the measured chirp mass the ``redshifted chirp mass" and the measured distance of the source the luminosity distance \citep{broadhurst18} while keeping other parameters the same. Thus, for simplicity, we will focus on the effect of acceleration and limit the discussion to the cases when cosmological and gravitational redshift can be neglected, i.e.,  $z\ll 1$. In particular, we assume $z_{\rm dop}\ll a_{//}f_{e}/2c\dot{f}_{e}$,  and simplify Eq.~(\ref{eq:acc1}) to be:
\begin{equation}
\dot{f}_{o}\approx \dot{f}_{e}+ f_{e}\frac{a_{//}}{c} \ .
\label{eq:acc2}
\end{equation}
The complete result for arbitrarily large redshift can be recovered by replacing all the $\mathcal{M}_{c}$ in the following discussion to be the redshifted chirp-mass $\mathcal{M}_{c}(1+z)$. 

Most of the proposed LISA mission strategies are to search for a signal using non-accelerating compact binary's GW templates. \citep[e.g.][]{Cornish+03,Littenberg+20} However, as discussed above we expect a degeneracy between $\mathcal{M}_{c}$ and $a_{//}$ since they both contribute to the frequency shift rate $\dot{f}_{o}$ in the leading order. For the case of a compact object binary with no or negligible eccentricity, only the 2nd harmonic of the GW signal dominates. In the LISA band, such kind of signal is a nearly-monochromatic sinusoidal wave with slowly increasing frequency. Since the waveform does not have other significant features except for $f_{o}$ and $\dot{f}_{o}$, neglecting the acceleration's contribution can lead to a biased estimation of the chirp mass. In particular, if one assumes $\dot{f}_{o}\approx \dot{f}_{e}+ f_{e}a_{//}/c$ (see Eq.~(\ref{eq:acc2})) as the intrinsic frequency shift rate $\dot{f}_{e}$ in template fitting, the resultant chirp mass will differ from the ``true'' chirp mass. This can be seen by combining Eq.~(\ref{eq:chirpmass-f}) and (\ref{eq:acc2}), which yields an observed chirp mass with the following expression:
\begin{align}
{\mathcal{M}_{o}}
& ={ \left(
\frac{5 c^5}{96 \pi^{8/3} G^{5/3}F(e) } \right) }^{3/5} (f^{j=2}_o)^{-11/5} (\dot{f}^{j=2}_o)^{3/5}\nonumber
\\& ={ \left(
\frac{5 c^5}{96 \pi^{8/3} G^{5/3}F(e) } \right) }^{3/5} (f^{j=2}_o)^{-11/5} (\dot{f}^{j=2}_e+a_{//}/c)^{3/5} \nonumber
\\ & \neq {\mathcal{M}_{c}} \ .\label{eq:chirpmass-observed}
\end{align}
Thus, there is a degeneracy between the binary's acceleration and mass in GW data analysis. In other words, the acceleration can significantly bias the chirp mass estimation if the compact binary has small or zero eccentricity and only the 2nd harmonic is detected. 
\subsection{Disentangling the Signatures of Accelerating Eccentric GW Sources - Analytical Approach}
\label{sec:disentangle}
For a GW source with peculiar acceleration, the chirp mass measurement without considering an accelerating template can lead to erroneous results (see examples above). However, when the source has non-negligible eccentricity and detectable multiple harmonics, the
GR precession creates a unique signature in the GW signal and disentangles the acceleration from the bias on the chirp mass. As a result, the accelerating GW signal will be different from any of the non-accelerating GW templates, thus can be identified.

Here we detailed the steps of this strategy. In particular, Section \ref{sec:eccToTheRescue} describes how the GR precession pattern of the eccentric compact binary leaves a signature on the GW waveform, thus differentiating the accelerating GW signal from the non-accelerating one. Section \ref{Analytical 2} estimates the critical acceleration for distinguishing accelerating GW sources in data analysis.
\begin{figure}[htbp]
\centering
\includegraphics[width=3.5in]{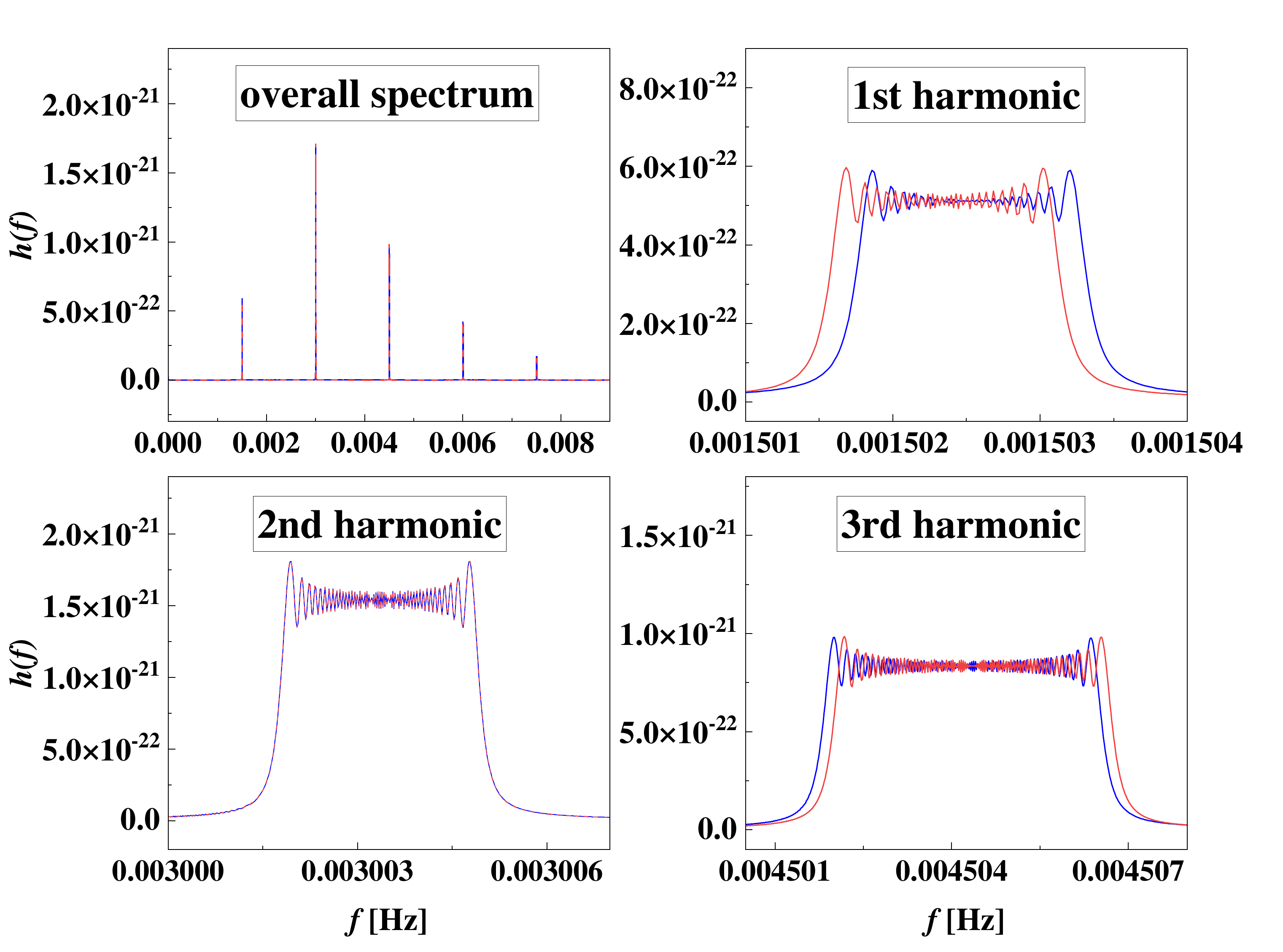}
\caption{{\bf The frequency spectrum of GW signals from two eccentric BBHs systems with the same chirp mass but different total mass.} For the first system (blue line) $m_{1}=8~\rm M_{\odot},$ $ m_{2}=40~\rm M_{\odot},$ $ e=0.3$; For the second system (red line) $m_{1}=m_{2}=16.867~\rm M_{\odot},$ $e=0.3$. Both of these systems have the same chirp mass $\mathcal M_{c}\sim 14.684~\rm M_{\odot}$, but as depicted the difference in total mass results in a different position of the harmonics. The initial period of radial motion for the first systems is $1.5 \rm mHz$, but its GW frequency of each harmonic is not the integer multiple of $1.5 \rm mHz$ because of the shift induced by GR precession ($\Delta f$). For the second system, its GR precession shifts the position of each harmonic by a different value. For illustration purposes, we adjust the initial radial frequency of the second system to be slightly different from $1.5 \rm mHz$, so that we compensate for the difference in $\Delta f$ and make its GW's $2$nd harmonic having the same initial frequency as the first system's.
 The observation duration is set to be two years. }
\label{fig:precession2}
\end{figure}
\subsubsection{Eccentricity as Means to Disentangle the Mass-Acceleration Degeneracy }\label{sec:eccToTheRescue}
The GR precession of an eccentric orbit shifts the periastron by an angle $\delta \phi_{0}$ for each period \cite{Misner+73}. Thus, the frequency of GW's n-th harmonic will split into a triplet $(nf-\Delta f, nf,nf+\Delta f)$ \citep{Seto+01,Moreno+95}, in which %
\begin{equation}
\Delta f= \frac{6(2\pi G)^{2/3}}{(1-e^{2})c^{2}}M_{\rm total}^{2/3}(P_{b})^{-5/3} \ ,\label{eq:precession}
\end{equation}
where $M_{\rm total}$ is the total mass of compact binary system, $P_{b}$ is the orbital period, and $f$ equals $1 /P_{b}$. For small eccentricity, coefficients of some terms in the triplet waveform are negligible, and the dominant component is $nf+\Delta f$. Therefore, the position of each GW's harmonic will be shifted by $\Delta f$ in the frequency domain. 

$\Delta f$ is independent of the harmonic number because it is a feature of the GR dynamic of the orbit.  Thus, all the harmonics will be shifted by the same value $\Delta f$. This feature allows us to disentangle $\Delta f$ and extract the total mass of the system, $M_{\rm total}$ \citep{Seto+01}. 

Figure~\ref{fig:precession2} demonstrates this feature of the GR precession pattern. In this figure, we adopt the x-model \citep{Hinder+10} to generate the GW signal from eccentric compact object binaries (see Section~\ref{NumericalA} for detailed information) and show the frequency spectrum of the GW signals from two different compact binary systems. They have the same chirp mass $\mathcal{M}_{c}$, which means that for each harmonic the time derivative of frequency (width of the peak in the frequency domain) is the same. But because of their different mass ratio, their total mass, $M_{\rm total}$, is different. Such difference in $M_{\rm total}$  yield a different  $\Delta f$ (see Eq.~(\ref{eq:precession})). Thus, even if we choose the special case when their GW's $2$nd harmonic has the same initial frequency/frequency shift rate (position/width), there will be a position displacement for their 1st and 3rd harmonics, which enables us to estimate their different $M_{\rm total}$. However, if their eccentricity is zero, only the 2nd harmonic is present, and we cannot distinguish such kind of precession pattern.

\begin{figure*}[htbp]
\centering
\includegraphics[width=7.5in]{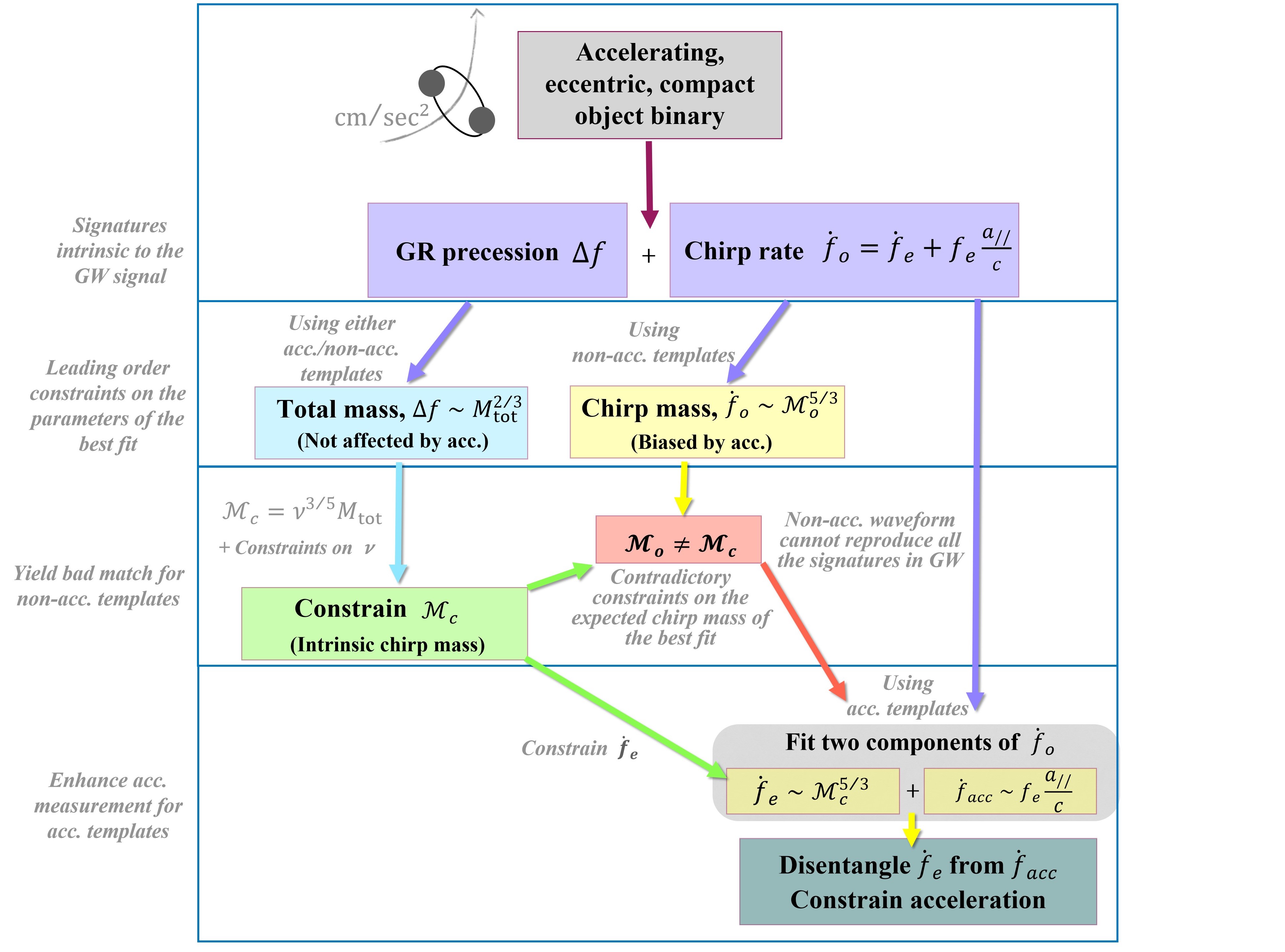}
\caption{ {\bf A flowchart of the proposed strategy, as the reason why the eccentricity can help us disentangle the acceleration in template fitting}.  The GW signal from accelerating eccentric sources has the signatures of GR precession ($\Delta f$) and chirp waveform ($\dot{f}_{o}$), which contribute differently to the result of template fitting. {\it Left branch} describes that $\Delta f$ primarily constrains the fitted total mass $M_{\rm tot}$, for either accelerating or non-accelerating templates. Since the total mass $M_{\rm tot}$ and the chirp mass $\mathcal{M}_{c}$ can be related to each other (See Eq.(\ref{eq:mc-mtotal0})), GR precession signatures in the waveform eventually put a constraint on the chirp mass $\mathcal{M}_{c}$ in template fitting. {\it Middle branch} shows the case when we use non-accelerating GW templates. In this case, the acceleration distorts the chirp rate ($\dot{f}_{o}\neq\dot{f}_{e}$) and yields an inaccurate chirp mass $\mathcal{M}_{o}$ in template fitting. However, the $\mathcal{M}_{c}$ from {\it left branch} (GR precession signature's constraint) is not affected by acceleration, thus will not agree with $\mathcal{M}_{o}$. Such a discrepancy between two waveform signatures means we can hardly find a good match in the non-accelerating templates family. It reminds us to include acceleration in the GW templates (i.e., shift to the {\it Right branch}). In the {\it Right branch}, we use GW templates with acceleration to fit both $\dot{f}_{e}$ and $\dot{f}_{\rm acc}$. Since the GR precession (from {\it left branch}) can constrain the intrinsic chirp rate $\dot{f}_{e}$, it disentangles $\dot{f}_{e}$ from $\dot{f}_{acc}$ and further help with improving the acceleration measurement accuracy.}
\label{fig:flowchart}
\end{figure*}

As shown in Section \ref{sec:AccGen}, the acceleration can yield a degeneracy between $\mathcal{M}_{c}$ and $a_{//}$ (see Eqs.~(\ref{eq:acc1}) and (\ref{eq:chirpmass-observed})). This phenomenon is due to the degeneracy in the level of observed frequency time derivative $\dot{f}_{o}$.   
However, estimating $M_{\rm total}$ from the $\Delta f$ caused by GR precession, only relies on the value of $f$, instead of its time derivative, (see Eq.~(\ref{eq:precession})). Thus, the total mass derived from the precession pattern is not affected by the degeneracy in the level of $\dot{f}$, therefore independent of the source's peculiar acceleration. 

Thus,  we propose that the acceleration signature can be disentangled from the GW signal for an eccentric source. Further, we suggest that a higher eccentricity can yield a higher sensitivity in the detection of acceleration. 

This strategy is described in  Figure \ref{fig:flowchart}. In particular, we firstly introduce the symmetric mass ratio that connects the chirp mass to the total mass in the following way: 
\begin{equation}
    \mathcal{M}_{c}=\nu^{3/5}M_{total}\ ,
    \label{eq:mc-mtotal0}
\end{equation}
 in which $\nu$ is the symmetric mass ratio:
\begin{equation}\label{eq:nu}
    \nu = \frac{m_1 m_2}{(m_1+m_2)^2} \ .
\end{equation}
The symmetric mass ratio always satisfies $\nu\leq 1/4$, and the value of $\nu$ is often constrained in GW template fitting. 

As illustrated in Figure \ref{fig:flowchart}, given an accelerating, eccentric compact object binary, its GW signal has the signatures of GR precession and GW emission. We can use templates (with/ without including the sources' peculiar acceleration) to fit the parameters of the source. In either case, the GR precession pattern ($\Delta f$) in the signal will constrain the total mass $M_{\rm total}$, which is not biased by the acceleration according to the reasons mentioned above. Thus,  Eq.~(\ref{eq:mc-mtotal0}), combined with the limits on $\nu$, places constraints on the intrinsic chirp mass $\mathcal{M}_{c}$ (see left branch in Figure \ref{fig:flowchart}). The limits on $\mathcal{M}_{c}$ can be used to constrain the intrinsic frequency shift rate $\dot{f}_{e}$ via Eq.~(\ref{eq:chirp mass}) (``Constrain $\dot{f}_e$'' sub-branch). 

On the other hand, the observed waveform has the information of both the frequency and its time derivative, i.e., $f_o$ and $\dot{f}_o$ (see the middle and right branch in Figure \ref{fig:flowchart}). Moreover, $\dot{f}_o$ has two components, one due to the intrinsic GW emission decay, and another caused by the peculiar acceleration. The former is $\dot{f}_{e}$, while the latter is $\dot{f}_{acc}\propto f_o\times a_{//}$ (see Eqs.~(\ref{eq:doppler2}) and (\ref{Eq:acc})). If we use non-accelerating GW templates to fit the signal (middle branch), the distorted $\dot{f}_o$ will result in a distorted chirp mass $\mathcal{M}_{o}$ (see Eq.~(\ref{eq:chirpmass-observed})). Therefore, the constraints of $\mathcal{M}_{c}$ we get from the GR precession signature in the left branch of Figure \ref{fig:flowchart} contradicts the $\mathcal{M}_{o}$ we get from the GW emission signature. In the template fitting, such a discrepancy means that two different signatures in the signal require contradictory parameters in the fitting result. In other words, in the parameter space of the non-accelerating templates there is no such a parameter set that can recover the signatures of both GR precession and GW emission. The bad match for non-accelerating templates may help us distinguish the acceleration. 

In the right branch, we use templates with acceleration to fit the signal. The constraints on $\dot{f}_{e}$ (from the left branch's GR precession signature) allow for disentangling the intrinsic frequency shift rate $\dot{f}_{e}$ from the observed quantity $\dot{f}_{o}$, thus yields constraints on the acceleration-induced frequency shift rate ($\dot{f}_{\rm acc} = \dot{f}_o-\dot{f}_e$). Therefore, the strategy in Figure \ref{fig:flowchart} overall explains why the eccentricity can help us distinguish (measure) the binary's acceleration when using GW templates without (with) acceleration.

We emphasize that here we adopt a simple case of one source signal. In practice, LISA may detect multiple sources simultaneously \citep[e.g.,][]{amaro17}, which can considerably add to the noise and confusion. However, the strategy we outlined here demonstrates that the GR precession signature in the signal disentangles the peculiar acceleration from the chirp-mass-induced frequency shift.
In other words, the eccentricity can cause the accelerating GW signal to differ from any non-accelerating GW template and constrains the acceleration-induced frequency shift rate, even when multiple sources are detected.

Below, we estimate the critical detection threshold that differentiate between the accelerating and non-accelerating GW sources.

\subsubsection{Analytical Analysis of the Detection Threshold}
\label{Analytical 2}
Let us consider a LISA-band GW waveform that is generated by an accelerating compact binary with non-negligible eccentricity (e.g., DWDs, initially $f_{e}^{j=2}=3\rm mHz$, $e>0.1$). In the cases that we consider, multiple harmonics of the waveform should be detected. For simplicity, below we neglect the constant redshift by assuming $z\ll 1$.

In this subsection we consider the implications of using non-accelerating GW templates to fit the GW signal. In particular, the GR precession pattern ($\Delta f$) constrains the intrinsic total mass of the system in template fitting. As outlined in Section \ref{sec:eccToTheRescue}, one can use Eq.~(\ref{eq:precession}) to extract the total mass, which is not affected by the extra $\dot{f}$ caused by $a_{//}$ since the measurement only depends on $f_{e}$ and $\Delta f$. However, the measured chirp mass $\mathcal{M}_{o}$ is different from the intrinsic chirp mass $\mathcal{M}_{c}$ for an accelerating source, because the acceleration changes the frequency shift rate $\dot{f}_o$ of the signal (i.e., Eq.~(\ref{eq:chirpmass-observed})).

We can relate the intrinsic $M_{\rm total}$ with the intrinsic $\mathcal{M}_{c}$ via the symmetry mass ratio $\nu$ (see Eq.(\ref{eq:mc-mtotal})). But as mentioned above, in the template fitting of accelerating GW sources, the $M_{\rm total}$ remains its intrinsic value while the $\mathcal{M}_o$ is biased. Thus, instead of being its intrinsic value ($\nu_{\rm int}$), the fitted symmetric mass ratio has to be biased ($\nu^{'}$) to make $\mathcal{M}_{o}$ and $M_{\rm total}$ agree with each other:
\begin{equation}
    \mathcal{M}_{o}=(\nu^{'})^{3/5}M_{\rm total} \ .
    \label{eq:mc-mtotal}
\end{equation}

In other words, the acceleration distorts the GW signal by changing the $\dot{f}_o$ (width of each harmonic in the frequency domain) but keeping the $\Delta f$ (the position displacement of each harmonic) unchanged. We can create a similar signature with non-accelerating waveforms by changing the symmetric mass ratio. Thus, sometimes it is still possible to find a non-accelerating GW template that fits the accelerating GW signal for the leading order. However, when the amplitude of acceleration increases, such degeneracy will eventually break.

In particular, since the symmetric mass ratio $\nu$ (and also $\nu^{'}$) has an upper bound of $0.25$, the observed chirp mass has a clear limit for a given total mass, i.e., $\mathcal{M}_{o}\leq0.25^{3/5}M_{\rm total}$. This limit can then be translated to a maximum acceleration $a_{\rm max}$ for which the observed chirp mass reaches $0.25^{3/5}M_{\rm total}$.  In other words, when the acceleration exceeds $a_{\rm max}$, $\mathcal{M}_{o}>0.25^{3/5}M_{\rm total}$, we cannot find a proper match in non-accelerating GW templates to fit both the $\mathcal{M}_{c}$ ($\dot{f}_o$) and the $M_{\rm total}$ ($\Delta f$). 

\begin{figure}[htbp]
\centering
\includegraphics[width=3.5in]{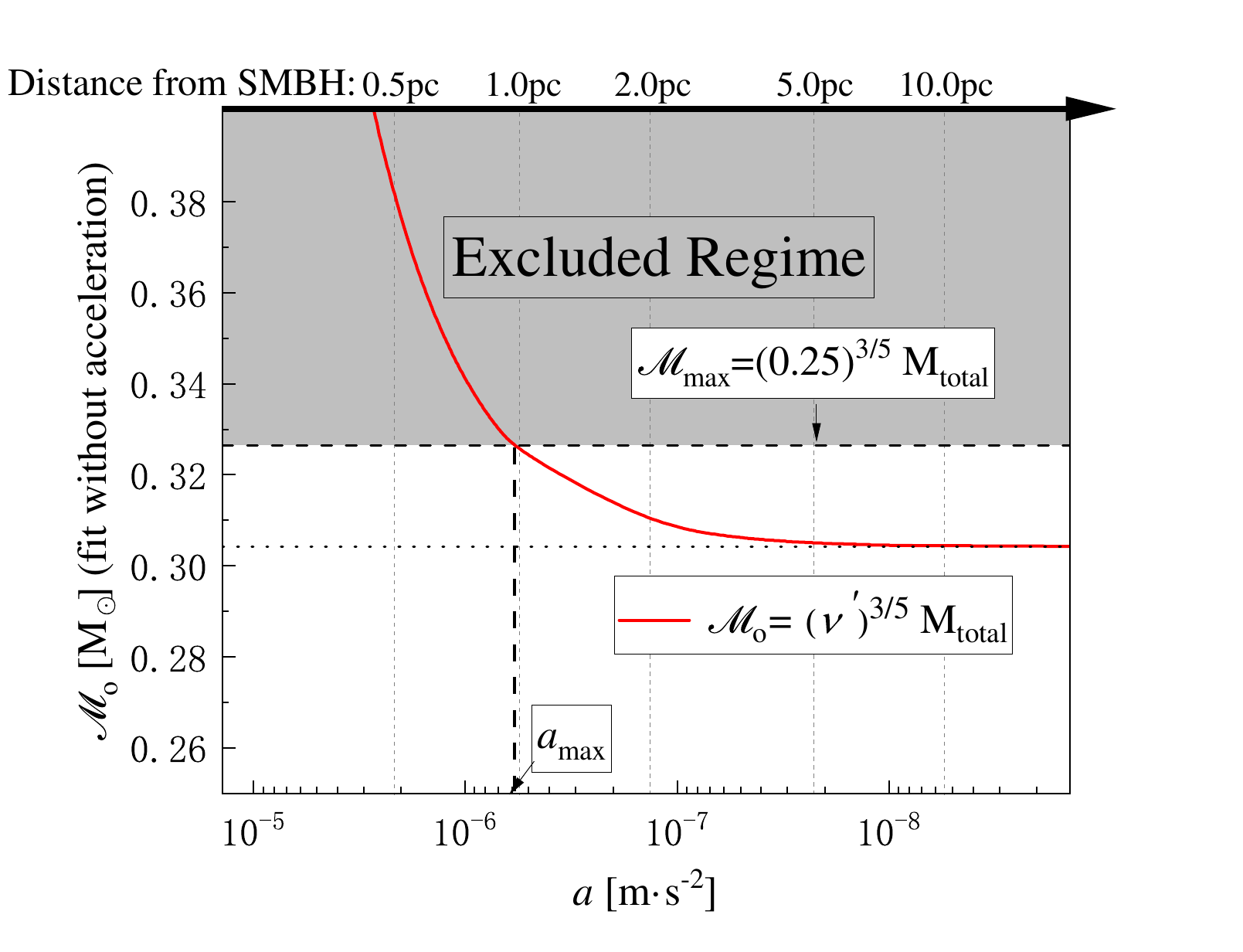}
\caption{{\bf The measured chirp mass $\mathcal{M}_{o}$ as a function of binary's acceleration.} We show a DWD system with  $m_{1}=0.25\rm {M_{\odot}},m_{2}=0.5\rm {M_{\odot}},$ $e=0.1, f_{e}^{j=2}=3\rm mHz$, on a  circular orbit around a SMBH with $M_{\rm SMBH}=4\times 10^{6}~\rm M_{\odot}$ at different distances. We choose the phase angle such that line-of-sight acceleration caused by the SMBH gravitational pull is maximized and pointing toward the observer. The chirp mass one measures using non-accelerating GW templates is depicted as a function of DWD's acceleration/ distance from the SMBH. As is shown in Section \ref{Analytical 2}, GR precession pattern will put a constraint on the maximum possible value of chirp mass: $\mathcal{M}_{c}\leqslant 0.25^{3/5}M_{\rm total}$. The figure shows that when the DWD system is close to the SMBH ($\sim 1\rm pc$), acceleration becomes large enough ($a>a_{\rm max}$) to make $\mathcal{M}_{o}$ exceed the upper bound $\mathcal{M}_{\rm max}$, enabling us to find the discrepancy between $\mathcal{M}_{o}$ and $M_{\rm total}$, thus distinguishing the acceleration. Note that the acceleration in this Figure increases to the left.}
\label{fig:mc-r}
\end{figure}

This relationship is depicted in Figure \ref{fig:mc-r}, where we consider an example of a DWD system at varying distances, and thus with varying acceleration, from a SMBH. As seen in the Figure, neglecting the contribution of acceleration in the template fitting can yield an observed chirp mass $\mathcal{M}_{o}$ which is larger than the allowed limit. Thus, implying that the binary exists in an accelerating environment. 

We now derive the expression of $a_{\rm max}$. As outlined in Section \ref{sec:AccGen}, comparing the difference between the intrinsic and observed GW frequency shift rate can be used to constrain the binary's peculiar acceleration.
 Using Eqs.~(\ref{eq:chirpmass-f}), ~(\ref{eq:acc1}), and ~(\ref{eq:chirpmass-observed}) we have:
\begin{equation}
\begin{aligned}
& \frac{a_{//}}{c}\approx \frac{\dot{f}_{o}-\dot{f}_{e}}{f_{e}}
\\ & =\frac{1}{f_{e}^{j=2}}\frac{96 \pi^{\frac{8}{3}}}{5}\left(\frac{G}{c^{3}}\right)^{5/3}(f_{e}^{j=2})^{11/3}F(e)\left(\mathcal{M}_{o}^{5/3}-\mathcal{M}_{c}^{5/3}\right) \ ,
\end{aligned}
\label{eq:doppler3}
\end{equation}
which describes the dependency of the observed acceleration on the difference between the observed ($\mathcal{M}_{o}$) and intrinsic ($\mathcal{M}_{c}$) chirp mass. 

As seen in Figure \ref{fig:mc-r}, the maximum acceleration is reached when the observed chirp mass $\mathcal{M}_{o}$ reaches it upper limit $\mathcal{M}_{\rm max}$: \begin{equation}
\mathcal{M}_{\rm max}=0.25^{3/5}M_{\rm total}=(\frac{0.25}{\nu_{\rm int}})^{3/5}\mathcal{M}_{c}\ .
\label{eq:cond1}
\end{equation}
Plug Eq.~(\ref{eq:cond1}) into Eq.~(\ref{eq:doppler3}), we can get:
\begin{equation}
\begin{aligned}
 \frac{a_{\rm max}}{c} & =\frac{F(e)}{f_{e}^{j=2}}\frac{96 \pi^{8/3}}{5}(\frac{G}{c^{3}})^{5/3}(f_{e}^{j=2})^{11/3}\mathcal{M}_{c}^{5/3}\left(\frac{0.25}{\nu_{\rm int}}-1\right)
\\ & = \frac{\dot{f}_{e}^{j=2}}{f_{e}^{j=2}}\left(\frac{0.25}{\nu_{\rm int}}-1\right)\ ,
\end{aligned}
\label{eq:doppler4}
\end{equation}
which is shown as the dashed vertical line in Figure \ref{fig:mc-r}. Thus, this maximum acceleration can help us estimate the detectability of the acceleration for eccentric GW sources. 

Additionally, in realistic observation there is noise. Thus, to distinguish the accelerating GW signal, the actual critical acceleration will need to be higher than the maximum acceleration $a_{\rm max}$ by a factor of $\delta a$:
\begin{equation}
a_{\rm crit}=\delta a+a_{\rm max} \ .
\label{eq:cond2}
\end{equation}

We can get $\delta a$ via the following reasoning: In general, to distinguish the acceleration, two conditions need to take place. First, the acceleration amplitude needs to be higher than $a_{\rm max}$, and second, the acceleration creates a significant difference in the waveform, which is larger than the noise. 

When $a<a_{\rm max}$ (right-hand side on Figure \ref{fig:mc-r}), the acceleration's signature can still be compensated for by changing the parameters of non-accelerating templates, and thus, harder to detect. On the other hand, for $a>a_{\rm max}$, we can no longer find a fit in the non-accelerating templates. However, due to the noise, the minimal acceleration that causes a fail fit in non-accelerating templates should be larger, see Eq.~(\ref{eq:cond2}). Therefore, the observed chirp mass will be larger than its original critical value when there is no noise. Using Eq.~(\ref{eq:doppler3}), once for $a_{\rm crit}$, and once for $a_{\rm max}$, we can express this higher chirp mass as follow: 
\begin{equation}
\begin{aligned}
\mathcal{M}^{'}= & \left(\mathcal{M}_{\rm max}^{5/3}+\frac{\delta a}{\frac{96}{5}\pi^{8/3}(f_{e}^{j=2})^{8/3}F(e)}\left(\frac{c^{3}}{G}\right)^{5/3}\right)^{3/5}
\\ & =0.25^{3/5}M_{\rm total}^{'}
\\ & >0.25^{3/5}M_{\rm total} \ .
\end{aligned}
\label{eq:new mc}
\end{equation}

Since $\nu^{'}$ has reached its maximum value of $0.25$, the total mass $M_{\rm total}^{'}$ in the template fitting has to be larger to be consistent with this new, and higher, chirp mass $\mathcal{M}^{'}$. Thus, the fitted template will have a larger shift for the frequency ($\Delta f^{'}$), due to the GR precession, than the shift that is observed ($\Delta f$). 
If the difference between $\Delta f^{'}$ and $\Delta f$ is larger than the frequency resolution of GW detector \citep{takahashi02}, we will have a clear signature of the acceleration. In other words, if 
\begin{equation}
\Delta f^{'} -\Delta f >  \frac{4\sqrt{3}}{\pi}\frac{\tau_{\rm obs}^{-1}}{\rm SNR} \ ,
\label{eq:cond3}
\end{equation}
where $\tau_{\rm obs}$ is the duration of observation, and SNR is the signal-to-noise ratio, the acceleration may result in a fail fit for a non-accelerating template. 

Using the Equations above, we can find the minimum magnitude of the acceleration ($a_{\rm crit}$) required to detect a signature of accelerating eccentric binary. In particular, we plug  Eq.~(\ref{eq:cond1})  into Eq.~(\ref{eq:new mc}) to get the $M_{\rm total}^{'}$ as a function of $M_{\rm total}$. Then we use Eq.~(\ref{eq:precession}) to get $\Delta f$ and $\Delta f^{'}$. We thus can use   Eq.~(\ref{eq:cond3}) to relate $\delta a$ with $M_{\rm total}$ and $\rm SNR$. Using the relationship between $a_{\rm crit}$ and $a_{\rm max}$ (Eq.~(\ref{eq:cond2})),  we finally get:
\begin{align}
\frac{a_{\rm crit}}{c} & \approx\frac{\dot{f}_{e}^{j=2}}{f_{e}^{j=2}}\frac{10\sqrt{3}(1-e^{2})}{3\pi^{5/3}}\frac{\tau_{\rm obs}^{-1}}{\rm{SNR}}(f_{e}^{j=2})^{-5/3}\left(\frac{G}{c^{3}}M_{\rm total}\right)^{-2/3}\nonumber
\\ & + \frac{\dot{f}_{e}^{j=2}}{f_{e}^{j=2}}\left(\frac{0.25}{\nu_{\rm int}}-1\right)\ .\label{eq: criteria}
\end{align}

This analytical expression of the critical acceleration is useful for estimating whether the GW signal from accelerating eccentric sources can be distinguished from non-accelerating GW templates, providing that more than one harmonic is detected with a signal-to-noise ratio above a certain SNR. In other words, if the acceleration of the GW source is larger than the critical acceleration in Eq.(\ref{eq: criteria}), we can distinguish between the accelerating signal and non-accelerating templates in observation. 

Additionally, this criteria is suitable for analyzing a wide range of sources, in particular,  double white dwarfs and double neutron stars (DNSs). However, the GW frequency of BBHs can change significantly during the observation because of their higher mass. This significant change of frequency contributes extra sensitivity in GW data analysis, thus limiting the potential usage of  Eq.~(\ref{eq: criteria}). Below, we will use numerical results to discuss the case of accelerating BBHs. We do emphasize that a dense place, such as the galactic center, is expected to host a high abundance of DWDs and potentially DNSs \citep[e.g.,][]{Stephan+19,Wang+21}. 

\subsection{Measuring the Acceleration of Eccentric Compact Binaries - Analytical Approach}
\label{sec:MeasureAcc}
Here we assume that an accelerating template for eccentric binaries is used in the data analysis. 
As mentioned before, the precession pattern can assist in distinguishing the eccentric binary's peculiar acceleration when we use non-accelerating GW templates to fit the signal. Thus, it is also possible that such a pattern can enhance the accuracy of acceleration measurement when we include the acceleration in the template. In this section, we  derive an analytical expression for the eccentric binary's measured acceleration and use it to estimate the accuracy of acceleration measurement for different kinds of compact binaries.

In particular, consider an accelerating eccentric compact binary system, for which the GR precession is detected so the total mass $M_{\rm total}$ can be extracted. As shown in Section \ref{sec:disentangle}, 
the measured frequency shift $\dot{f}_{o}$ has two parts: intrinsic frequency shift $\dot f_{e}$ caused by GW radiation, and an extra term caused by the acceleration. In order to determine the extra term and measure the acceleration, we need to subtract $\dot f_{e}$ from $\dot{f}_{o}$ (see Eq.~(\ref{eq:doppler3})). However, $\dot f_{e}$ is a function of the intrinsic chirp mass $\mathcal{M}_{c}$, while we can only extract the $M_{\rm total}$ from the precession pattern. In fact, $\mathcal{M}_{c}$ depends on both $M_{\rm total}$ and $\nu$ (see Eq.~(\ref{eq:mc-mtotal0})). Thus, an eccentric binary with a total mass measured from its GR precession will still have uncertainty on its acceleration measurement, which is coupled to the symmetric mass ratio $\nu$. 

The acceleration has straightforward dependence on the symmetric mass ratio $\nu$. Thus, for a range of possible $\nu$,  
the line of sight acceleration $a_{//}$ can be expressed as:
\begin{equation}\label{eq:acc-measurement}\frac{a_{//}}{c}\approx\frac{\dot{f}^{j=2}_{o}}{f^{j=2}_{o}}-\frac{96\pi}{5}
(\Delta f_{o})^{5/2}(f^{j=2}_{o})^{-3/2}\left(\frac{1-e^{2}}{3}\right)^{5/2}F(e)\nu \ ,
\end{equation}
$f_{o}^{j=2}$  ($\dot{f}_{o}^{j=2})$ is the observed frequency (the observed frequency shift) of GW's second harmonic; $\Delta f_{o}$ is the observed frequency difference caused by precession; $e$ is the eccentricity of binary, which can be got by comparing the relative amplitude of each harmonic. Eq.~(\ref{eq:acc-measurement}) was derived by using Eqs.~(\ref{eq:chirp mass}),  (\ref{eq:doppler2}) and (\ref{eq:precession}), and substitute $\mathcal{M}_{c}$ with $\nu^{3/5} M_{\rm total}$. We have also assumed that the redshift of the binary is small.

All the other quantities on the right side of Eq.~(\ref{eq:acc-measurement}) can be directly found by measuring the frequency, frequency evolution, and amplitude for different harmonics of the GW waveform, except for $\nu$. Thus, the only unknown, in this case, is $\nu$, for considering the leading order features of the GW signal. However, since $\nu$ will affect the frequency evolution in higher order, it can be constrained using template fitting in the data analysis. The error of the measured $\nu$ and its influence on acceleration measurement can be different, depending on the specific parameters of the GW source. Thus, we will adopt DWDs/BBHs as examples in the following discussion.

DWDs typically have a smaller chirp mass than BBHs,  which means their intrinsic frequency shift rate of GW (``chirp rate" hereafter) is smaller. In other words, DWDs behave like monochromatic GW sources. As mentioned above, the lack of intrinsic frequency evolution renders an under-constrained $\nu$ for DWDs. However, since $\Delta f\propto M_{\rm total}^{2/3}$, the coefficient before $\nu$ is also smaller for DWDs in Eq.~(\ref{eq:acc-measurement}). Therefore acceleration measurement of DWDs has a weaker dependency on $\nu$ compared to the BBHs case.  Thus, even if DWDs' $\nu$ is highly uncertain, we can still establish an accurate acceleration measurement in the observation. 
\begin{figure}
  \centering
\includegraphics[width=\columnwidth]{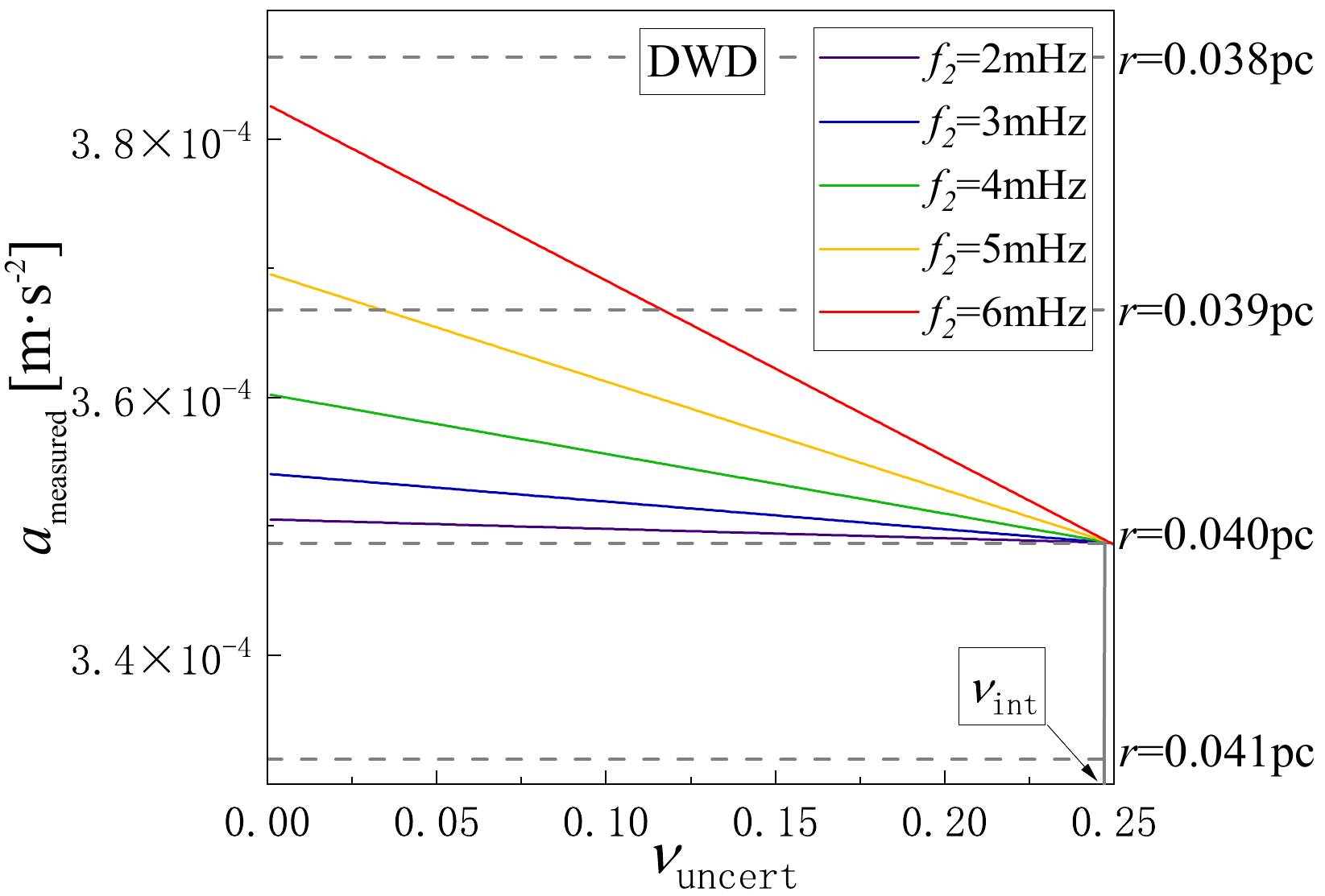}\\
\includegraphics[width=\columnwidth]{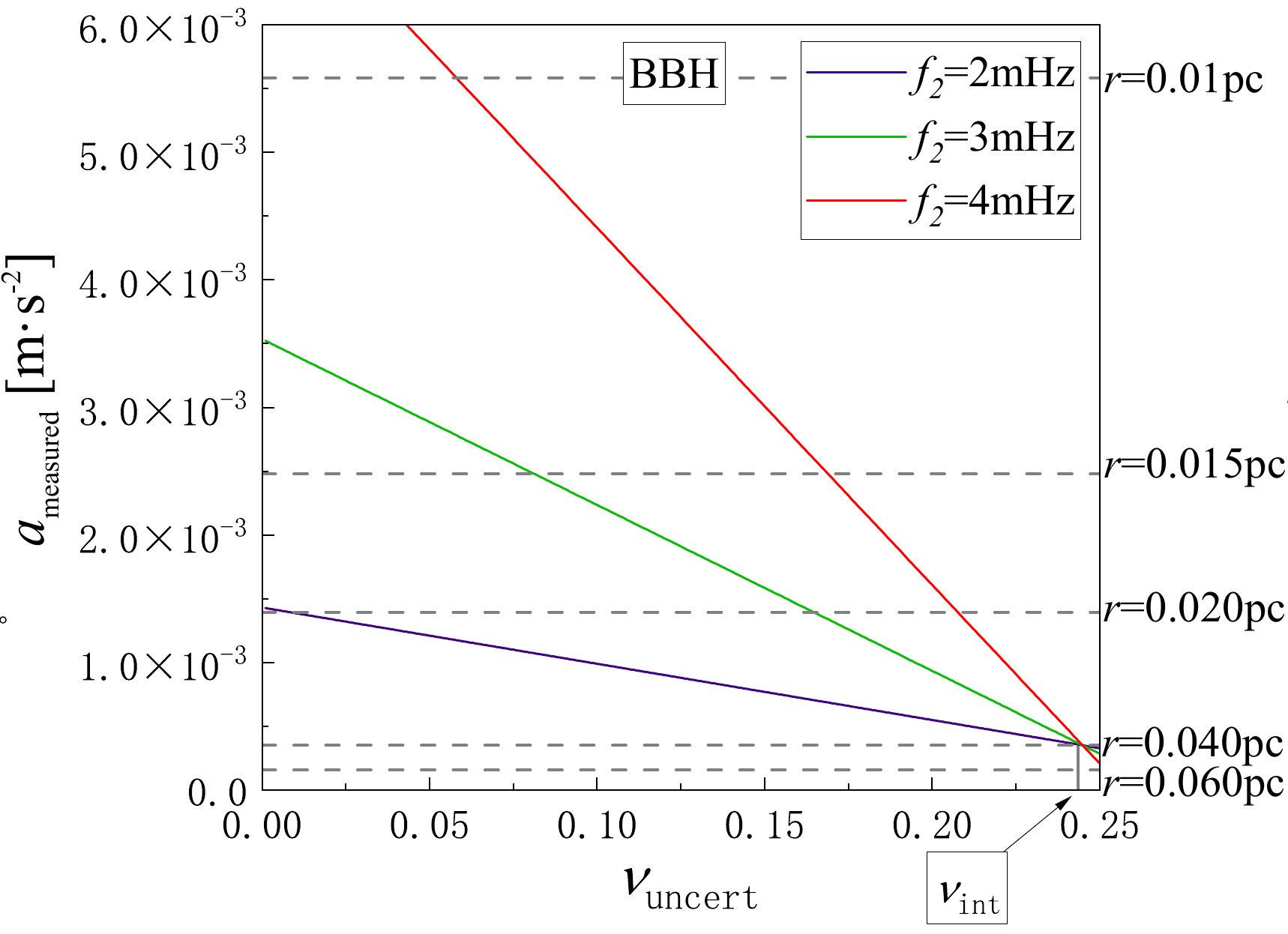}\\
\hspace{-1.1cm}
\includegraphics[width=0.9\columnwidth]{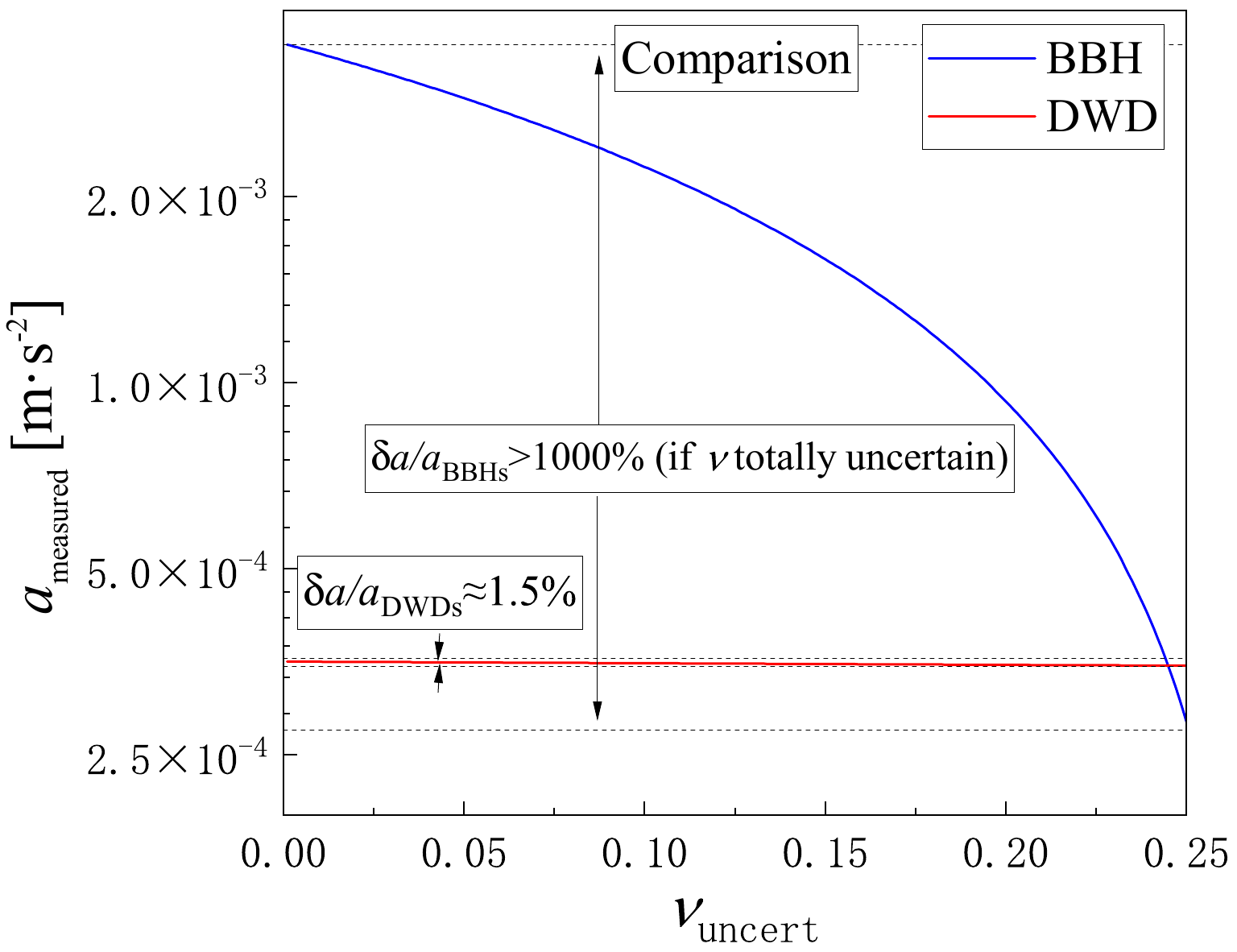}
\vspace{-0.4cm}
  \caption{{\bf The measured acceleration as a function of the uncertainty in measured symmetric mass ratio $\nu_{\rm uncert}$} 
 Here we place a binary at $0.04$~pc from a $4\times10^{6}\rm M_{\odot}$ SMBH and choose the phase angle so that the line-of-sight acceleration is maximized. 
{\it Upper Panel} shows a binary WD composed of: $m_1=0.35$~M$_{\odot}$, $m_2=0.4$~M$_{\odot}$, $e=0.1$, with the frequency of second harmonic $f_{2}=2,3,4,5$ and $6$~mHz, bottom to top. 
{\it Middle Panel} depicts a BH binary composed of: $m_1=15$~M$_{\odot}$, $m_2=20$~M$_{\odot}$, $e=0.1$. 
{\it Bottom Panel} is the same as the top and middle panel, but shows the comparison between DWDs and BBHs for $f_{2}=3\rm mHz$ to scale.
The solid lines represent the measured acceleration as a function of measured $\nu$, (see Eq.~(\ref{eq:acc-measurement})). The dashed lines correspond to the magnitude of peculiar acceleration at the labeled distances from the SMBH.  As noted in the text, the measurement of the symmetric mass ratio can be uncertain, depicted here in the x-axis by  $\nu_{\rm uncert}$ (all the possible values of $\nu$).The intrinsic symmetric mass ratio $\nu_{\rm int}$ of the binary is marked on the plot for reference.}
\label{fig:dwd-ana}
\end{figure}

This behavior is depicted in the upper panel of Figure \ref{fig:dwd-ana}, where we show the acceleration as a function of the symmetric mass ratio $\nu$ (see Eq.~(\ref{eq:acc-measurement})). To highlight that the symmetric mass ratio is not constrained, we denote it in the Figure as ``uncert,'' i.e., $\nu_{\rm uncert}$. As seen in the Figure, because the DWDs' chirp rate ($\dot{f}_{e}$) is relatively small, the frequency shift we observe mostly comes from the source's peculiar acceleration (``$f_{e}a_{//}/c$" in Eq.~(\ref{eq:acc2})). Thus, the DWDs have a small error of acceleration measurement and a clear signature of the accelerating environment. Even for a highly under-constrained symmetric mass ratio, the accuracy of acceleration measurement is still promising, about $0.5-10\%$ from the lowest to highest frequency (compared to the dashed line in the Figure, and also see the bottom panel).

On the other hand, BBHs' chirp rate is relatively large, enabling us to determine $\nu$ better. But the acceleration measurement of BBHs highly depends on $\nu$ because of their large total mass. Therefore, an inaccurate $\nu$ in the GW data analysis of BBHs can induce an extremely large error in the acceleration measurement. For example, in the middle and lower panel of Fig.~\ref{fig:dwd-ana}, error of acceleration measurement is over $100\%$ for a BBH system with the frequency of the $2$nd harmonic $2-4$~mHz, which is not acceptable in the observation. 

To conclude, DWDs' acceleration measurement accuracy is promising even if there is no constraint on the symmetric mass ratio $\nu$ in observation. Thus, it can be estimated analytically. However, the estimation of BBHs' acceleration measurement error requires the numerical GW template fitting, in which we analyze the higher-order frequency evolution and determine the accuracy of $\nu$ measurement. We address this issue for the case of BBHs using the numerical method in Section \ref{egB}.

For a binary at the center of a galaxy, other physical processes may also affect its evolution. In particular, the Eccentric Kozai-Lidov \citep[EKL, e.g.,][]{kozai62,lidov62,Naoz16}, can excite eccentricities and enhance the compact objects merger rate \citep[e.g.,][]{Hoang+18,Stephan+19,Wang+21}.  
However, in our case, we focus on binaries that will appear in the LISA band (i.e., in the mHz frequencies), in other words, hard binaries, where the EKL timescale is very long. For example,  a $3$~mHz GW  signal for a $35$~M$_\odot$ circular binary system corresponds to a binary separation of $\sim0.0025\rm au$.   Specifically, 
\begin{equation}
\begin{aligned}
&T_{\rm{LK}} \sim \frac{P_{\rm SMBH}^2}{P_{b}} (1-e^2)^{3/2}
\\ & \sim 2\times 10^{10} {\rm{yr}} 
	\\ & \times \left(\frac{m_{\rm SMBH}}{4 \times 10^6 M_{\odot}}\right)^{-1}
	\left(\frac{f_e^{j=2}}{3 \rm{mHz}}\right) 
	\left(\frac{r}{0.1 \rm{pc}}\right)^{3} (1-e^2)^{3/2}\ ,
	\label{eq:TKL}
\end{aligned}
\end{equation}
where $P_{\rm SMBH}$ is the period of the binary around the SMBH. Therefore, for these hard binaries, EKL effects can be neglected \footnote{Note that for the binaries that are in close proximity to the SMBH, the EKL mechanism can still result in eccentricity excitations for hard binaries that are potentially detectable in LISA \citep{Hoang+19}.}. Similarly, the interactions with flying neighbors will only tend to harden the binary \citep[e.g.,][]{heggie75}. Moreover, two-body relaxation may allow the binary to migrate in a diffusive manner over time, however, this process is also much longer than the observational timescale and thus can be neglected \citep[e.g.,][]{Sari+19,Rose+20}. 

We emphasize that the peculiar acceleration and the velocity of the binary are still relatively small here ($\sim 10^{2} ~km\cdot s^{-1}$ in this example), thus do not require any special relativity corrections. 
Additionally, when measuring the acceleration of the eccentric compact binaries, other factors can contribute differently to the precession pattern in the GW signal (e.g., spin effect and tidal effect, the latter can be important for DWDs and BNSs \citep{Lau+22}). 
The full inclusion of these effects is beyond the scope of this paper. Moreover, unlike the previous studies of circular binary's acceleration  \citep[e.g.,]{Seto+08,Robson+18,Tamanini+20}, the acceleration of eccentric compact binaries is measurable even when they are distant from the tertiary, up to $\sim 1\rm pc$ for a BBH system around a $4\times 10^{6}\rm{M}_{\odot}$ SMBH, with the orbital period ($\sim 10^{3}$~yrs) much longer than the observation duration.

\section{Numerical Approach}
\label{Numerical}
\subsection{Waveform Template }
\label{NumericalA}
For the generation of the accelerating eccentric compact binaries' GW templates, we adopt the x-model developed in Ref.~\citep{Hinder+10} and add the effect of peculiar acceleration by including the time-dependent doppler effect in waveform modulation.

The x-model is a time-domain, post-Newtonian (pN)-based waveform family, which captures all the critical features that eccentricity introduces to non-spinning binary 
\citep{Huerta+14}. The binary orbit is given in the
Keplerian parameterization to $3$~pN order and the conservative evolution is given to $3$~pN order as well. In the x-model, the loss of energy and angular momentum is mapped to the change of orbital eccentricity $e$ and the pN expansion parameter $x\equiv (\omega M_{\rm total})^{2/3}$, in which $\omega$ is the mean Keplerian orbital frequency. These two parameters are evolved according to $2$~pN equations. 

The equations in the x-model can be numerically solved, and thus can be used to generate eccentric GW templates in the time domain. The results have been validated against numerical relativity for the case of equal mass BBHs, $e=0.1$, 21 circles before the merger. The x-model also reduces to some well-studied template families in the GW data analysis for the zero-eccentricity case \citep{Brown+10}. 

We remind the reader that here, as a proof of concept, we focus on the acceleration-induced, time-dependent doppler effect. 
For demonstration purposes, in the following examples, the tertiary is placed in a circular orbit, and the phase angle is set to make the line-of-sight acceleration towards the detector maximized. However, we emphasize that different phase angles can lead to different signatures of acceleration in the GW waveform, and the specific value of acceleration measurement accuracy also varies.

\subsection{Matched Filtering and Criteria for Distinguishing the Acceleration}
\label{NumericalB}
To numerically verify whether the acceleration can be identified or not, we need to introduce some standard definitions in the GW data analysis. Here we use a numerical technique called ``matched filtering”, which is commonly adopted in the GW data analysis to estimate the GW sources' parameters \citep{Dhurandhar+94, thorne1987300, Finn92,Cutler+94}. 

For our purpose, we use this technique to quantify the similarity
between two different waveforms. In particular, we will first transform the GW signal from an accelerating eccentric compact binary into the frequency domain (denoted as $h_{1}$ hereafter), then use a set of non-accelerating GW templates, $h_{2}$, to fit the detected signal $h_{1}$. We treat the waveforms as vectors in a Hilbert space \citep{Helstrom+13}, and define the noise-weighted inner product between $h_{1}$ and $h_{2}$ as: 
\begin{equation}
\left\langle h_{1} \mid h_{2}\right\rangle=2 \int_{0}^{\infty} \frac{\tilde{h}_{1}(f) \tilde{h}_{2}^{*}(f)+\tilde{h}_{1}^{*}(f) \tilde{h}_{2}(f)}{S_{\mathrm{n}}(f)} \mathrm{d} f \ ,
\label{eq:innerproduct}
\end{equation}
in which $\tilde{h}_j$ (with $j=1,2$) means a Fourier transformation
of the waveform, the star stands for the complex conjugate, and $S_{n}(f)$ is the one-sided noise power spectral density of LISA \citep{thorne_1987,Klein+16}.

The similarity between $h_{1}$ and $h_{2}$ can be estimated by the Match between them (e.g., see Section IV of Ref.\citep{Afle+18} for a summary). First, define the normalized overlap as: 
\begin{equation}
\mathcal{O}(h_{1}, h_{2})=\langle\hat{h}_{1}| \hat{h}_{2}\rangle=\frac{\langle h_{1}| h_{2}\rangle}{\sqrt{\langle h_{1}| h_{1}\rangle\langle h_{2}| h_{2}\rangle}} \ .
\end{equation}
The Match between $h_{1}$ and $h_{2}$ is defined as the maximized overlap over the signal's arrival time $t_{0}$ and phase $\Phi_{0}$: 
\begin{equation}
\rm{Match}(h_{1}, h_{2})=\max _{t_0, \Phi_0} \mathcal{O}(h_{1}, h_{2})\ .
\end{equation}
For a given $h_{1}$, we will generate a set of non-accelerating waveforms $h_{2}$, with different values of template parameters, to match $h_{1}$. The maximized Match over all the template parameters of $h_{2}$ is called the fitting factor (FF):
\begin{equation}
\rm {F F}=\max_{\boldsymbol{\lambda}}  \{\rm{Match}(h_{1}, h_{2})\} \ ,
\label{eq:FF}
\end{equation}
in which $\boldsymbol{\lambda}$ represents the parameters of the GW template. (e.g., chirp mass $\mathcal{M}_{c}$, symmetric mass ratio $\nu$, etc.)

In general, the absolute value of FF varies from $0$ to $1$, and a perfect match between two waveforms would give $\rm{FF} = 1$.  

Additionally, we need to consider the strength of the signal, which is quantified by the signal-to-noise ratio ($\rm SNR$). For a given $\rm SNR$ level, there should exist a region in the parameter space where we cannot tell between two similar GW templates because of the noise in observation. The stronger the signal is (i.e., larger SNR), the smaller the confusion region will be. Since $\rm FF$ evaluates the similarity between two GW waveforms, we can compare the $\rm FF$ with $\rm SNR$ and derive a rule-of-thumb criterion for the two waveforms to be indistinguishable in the GW data analysis: \citep{lindblom08,2017PhRvD..95j4004C,2002ApJ...575.1030T,Cutler+07}: 
\begin{equation}
	{\rm FF}> 1-(D-1)/(2~\rm{SNR}^2)\ ,
	\label{eq:criteriaFF}
\end{equation}
in which $D$ represents the dimension of parameter space where the matched filtering is carried out ($\sim10$ for LISA-band sources, depending on the template. In our case, the parameters include mass for the components of binary $m_{1},m_{2}$; eccentricity $e$; orbital frequency $f$; the distance of the source $d$; the angular position of the source $\phi_{S},\theta_{S}$; orientation of the source $\phi_{L},\theta_{L}$; and the line-of-sight acceleration $a_{//}$).

To conclude, we generate a set of non-accelerating GW waveforms $h_{2}$, with different parameters, then compute the match between an accelerating GW signal, $h_{1}$, with $h_{2}$, using Eq~(\ref{eq:FF}). Denoting the value of the optimal match as $\rm FF$, it represents how similar the accelerating GW signal can be to the non-accelerating waveforms. Thus, in the observation, if $\rm{FF}$ is below the threshold given by the right side of Eq~(\ref{eq:criteriaFF}), it means the accelerating GW source's signal is distinguishable from any non-accelerating GW templates, i.e., strong enough for us to find the signature of acceleration in GW data analysis.

For demonstration purposes, in this paper, we will plot the quantity $\rm 1-FF$ instead of $\rm FF$ to highlight the difference between the two waveforms. Since $\rm FF$ is the maximized Match over all the template parameters, $\rm 1-FF$ represents the minimized Mismatch ($\rm 1-Match$) between a GW signal and its optimal fit. Similar to the criteria mentioned above, if this minimized mismatch becomes large enough, i.e., $\rm {1-FF}>(D-1)/(2~\rm{SNR}^2)$, there will be a distinguishable difference between the accelerating and non-accelerating waveforms.

\subsection{Fisher Matrix Analysis and Measurement Accuracy}
\label{NumericalC}
Adopting the GW templates with acceleration, we can measure the value of compact binaries' peculiar acceleration in the data analysis. In this case, the accuracy of acceleration measurement can be evaluated using the Fisher Matrix Analysis \citep{Coe+09,Cutler+94}, which is commonly used as a linearized estimation for the measurement error in the high $\rm SNR$ limit.

Representing the parameters of a GW source as a vector $\boldsymbol{\lambda}$, the GW waveform $h$ can be expressed as $h(\boldsymbol{\lambda})$. The Fisher Matrix is defined as:
\begin{equation}
    F_{ij} = \left\langle\frac{\partial h(\boldsymbol{\lambda})}{\partial \lambda_i},\frac{\partial h(\boldsymbol{\lambda})}{\partial \lambda_j}\right\rangle.
\end{equation}
We define $C$ as the inverse of the Fisher matrix, $C = F^{-1}$. It approximates the sample covariance matrix of the Bayesian posterior distribution for the GW source's parameters. In other words, we can use the following equation to estimate the error of parameter measurement in our work:
\begin{equation}
\delta \lambda_{i}= \sqrt{\left\langle\left(\Delta \lambda_{i}\right)^{2}\right\rangle}=\sqrt{C_{i i}}
\end{equation}
In a nutshell, when estimating the accuracy of acceleration measurement, the waveform templates are generated numerically according to the method mentioned in Section~\ref{NumericalA}, including the peculiar acceleration in the parameter set. We then adopt Fisher Matrix Analysis to calculate the error of the acceleration measurement for different kinds of GW sources.
\section{Examples and Results}
\label{examples}

Here we provide several examples to highlight the usefulness and application of the methods in Section~\ref{Numerical}. We emphasize that the measurement accuracy of acceleration is only valid if we distinguish the accelerating GW sources from the non-accelerating ones. Thus, adopting non-accelerating GW templates in search of GW signals can suppress us from identifying some accelerating GW sources. As mentioned in Section~\ref{NumericalC} and \ref{NumericalB}, these sources can have good acceleration measurement accuracy but are misidentified because they fail to meet the criteria for distinguishing the acceleration. However, we can overcome this drawback by using the accelerating GW templates when searching for the GW signal. 

In other words, the criteria in Section~\ref{NumericalB} and the examples in this Section~\ref{egA} are for the cases when we merely use non-accelerating templates in search of GW signals, thus being more stringent. However, the analysis of Section~\ref{NumericalC} and examples of Section~\ref{egB} assume that the acceleration has been identified and included in the GW template fitting. Thus, their results represent the maximum capability of LISA to measure the acceleration of eccentric compact binaries.

\subsection{Distinguishing the Acceleration - Numerical Approach}
\label{egA}
\begin{figure}[htbp]
\centering
\includegraphics[width=3.5in]{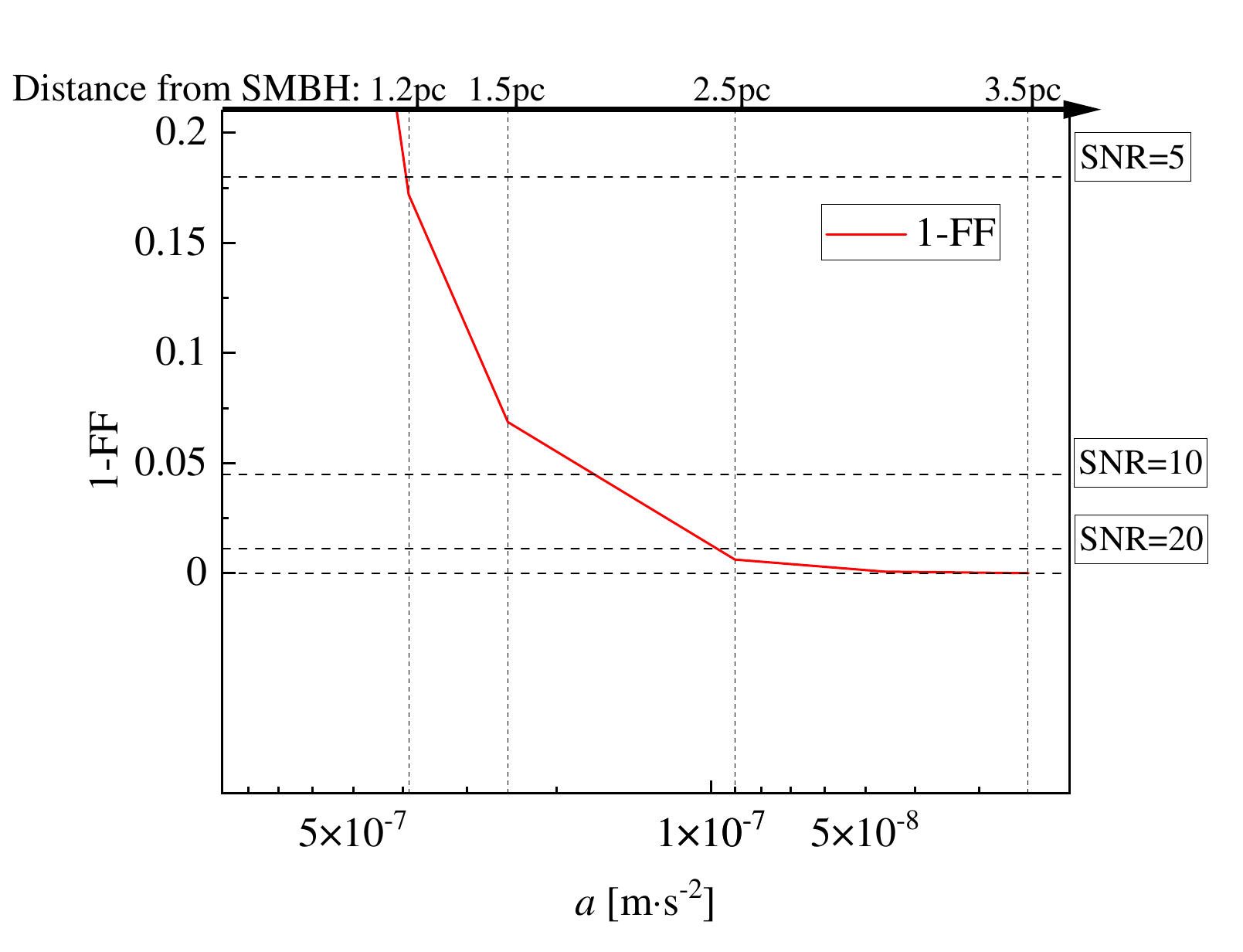}
\caption{{\bf The mismatch between an accelerating DWD's GW and its optimal fit within non-accelerating templates ($\rm 1-FF$), as a function of the DWDs' peculiar acceleration.} We show a DWD system with  $m_{1}=0.35\rm {M_{\odot}},m_{2}=0.4\rm {M_{\odot}},$ $e=0.1, f_{e}^{j=2}=3\rm mHz$, on a  circular orbit around a SMBH with $M_{\rm SMBH}=4\times 10^{6}~\rm M_{\odot}$ at different distances. The observation duration is set to be 4 years. Adopting the proposed methodology of future observations, we apply matched filtering method \citep{Finn92,Cutler+94} to find the best fit in templates without considering acceleration. $1-{\rm FF}$ is plotted as a function of DWD's 
acceleration/ distance from the SMBH. The black lines mark the critical value of $1-\rm{FF}$ for the detection of acceleration, providing different SNR in observation (see Eq.~(\ref{eq:criteriaFF}), $D=10$ for this example). Note that the acceleration in this Figure increases to the left.}
\label{fig:mismatch_dwd}
\end{figure}
In Section~\ref{Analytical 2}, we derived the criteria for distinguishing the acceleration (Eq.~(\ref{eq: criteria})). This equation is valid when the mass of a compact binary is small and the curvature of frequency evolution is hard to measure (e.g., DWDs). Here we compare this Equation with numerical results as well as expand it to other cases, such as BBHs. 

\begin{figure}[htbp]
\centering
\includegraphics[width=3.5in]{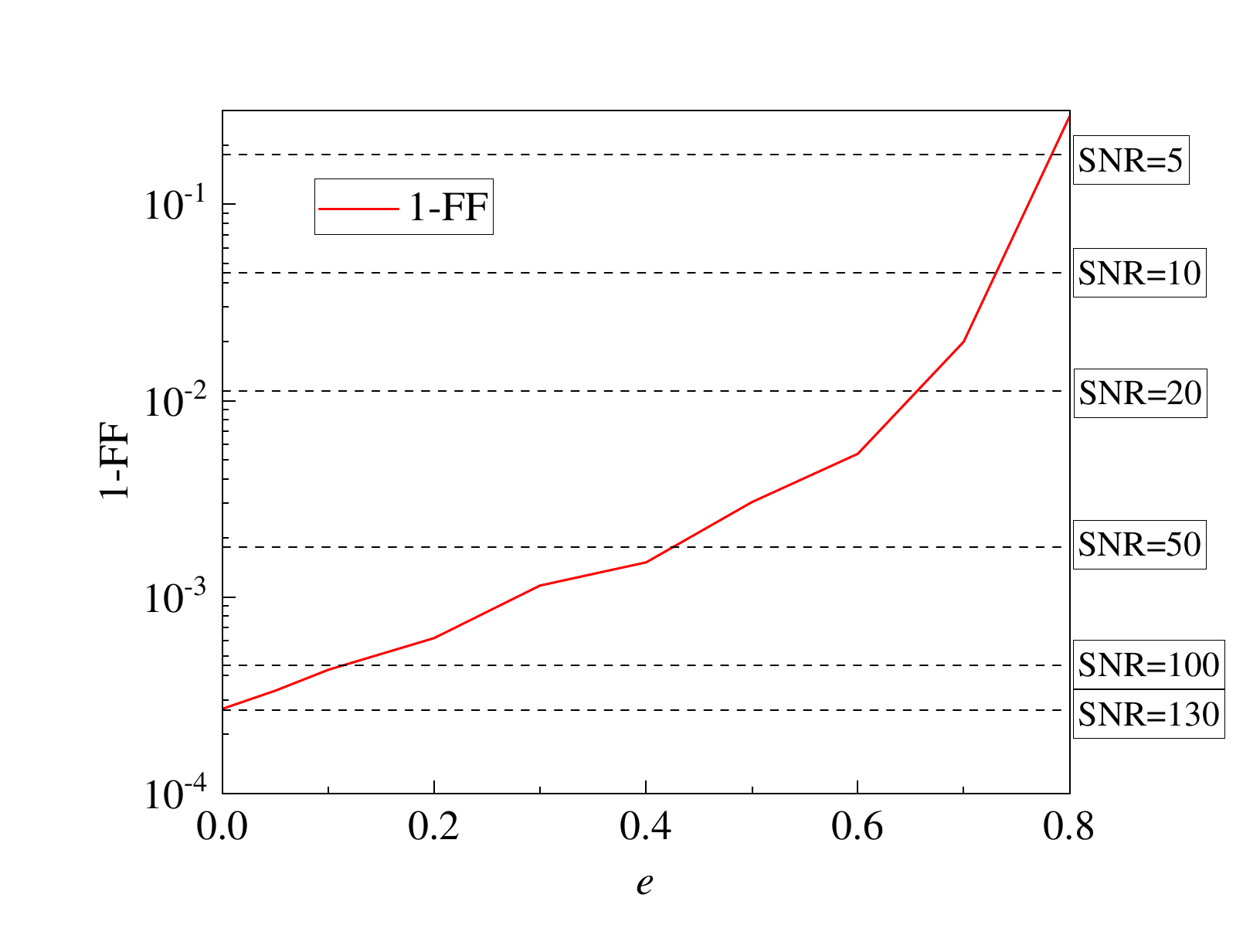}
\caption{{\bf The mismatch between an accelerating BBH's GW and its optimal fit within non-accelerating templates ($\rm 1-FF$), as a function of the BBH's eccentricity.} Here we place a BBH system with $m_{1}=15 {\rm{M}}_{\odot}, m_{2}=20 {\rm{M}}_{\odot}, f_{e}^{j=2}=3\rm mHz$ at $0.1$~pc from a $4\times10^{6}\rm M_{\odot}$ SMBH, set the observational time $\tau_{{\rm obs}}=4\rm yrs$. Following the same steps as Figure~\ref{fig:mismatch_dwd}, we apply matched filtering method to find the best fit in templates without considering acceleration and plot $\rm 1-FF$ as a function of BBH's eccentricity. As is shown in the figure, $\rm 1-FF$ becomes larger and the acceleration is more likely to be detected for higher eccentricity.}
\label{fig:mismatch_bbh}
\end{figure}

The first example is depicted in Figure~\ref{fig:mismatch_dwd}, where we consider a DWD system orbiting around a SMBH. Note that we expect a high abundance of DWDs to reside in the vicinity of our own SMBH \citep{Stephan+19,Wang+21}. This Figure shows $\rm 1-FF$, the mismatch between the accelerating signal and its optimal fit within non-accelerating templates, as a function of the source's acceleration. As shown in the Figure, larger acceleration values yield a larger mismatch between the accelerating signal and non-accelerating templates. Comparing Figure~\ref{fig:mismatch_dwd} with Figure~\ref{fig:mc-r}, while the y-axis is different in these two Figures, we recover a similar trend as in the analytical estimation. 

In particular, when $\rm 1-FF$ exceeds each dashed black line in Figure~\ref{fig:mismatch_dwd}, we can distinguish the acceleration of the GW source for the corresponding $\rm SNR$ level. For the configuration in this Figure, the mismatch becomes large enough for us to distinguish the acceleration when the DWD's distance to the SMBH is $2.4\rm{pc},1.9\rm{pc},1.2\rm{pc}$ (for the $\rm{SNR}=20,10,5$, respectively). This result is consistent with the analytical estimation in Figure~\ref{fig:mc-r}, where the chirp mass measurement significantly deviates from the intrinsic chirp mass and exceeds the upper bound given by the precession pattern at around $1\sim2 \rm pc$. 

Figure~\ref{fig:mismatch_dwd} highlights that the critical distance for distinguishing the acceleration is much larger than the estimation of former works \citep{Robson+18,Tamanini+20}. This difference is because we allowed for a non-negligible, though still small, eccentricity, $e=0.1$, in our example.
\begin{figure}[htbp]
\centering
\includegraphics[width=3.5in]{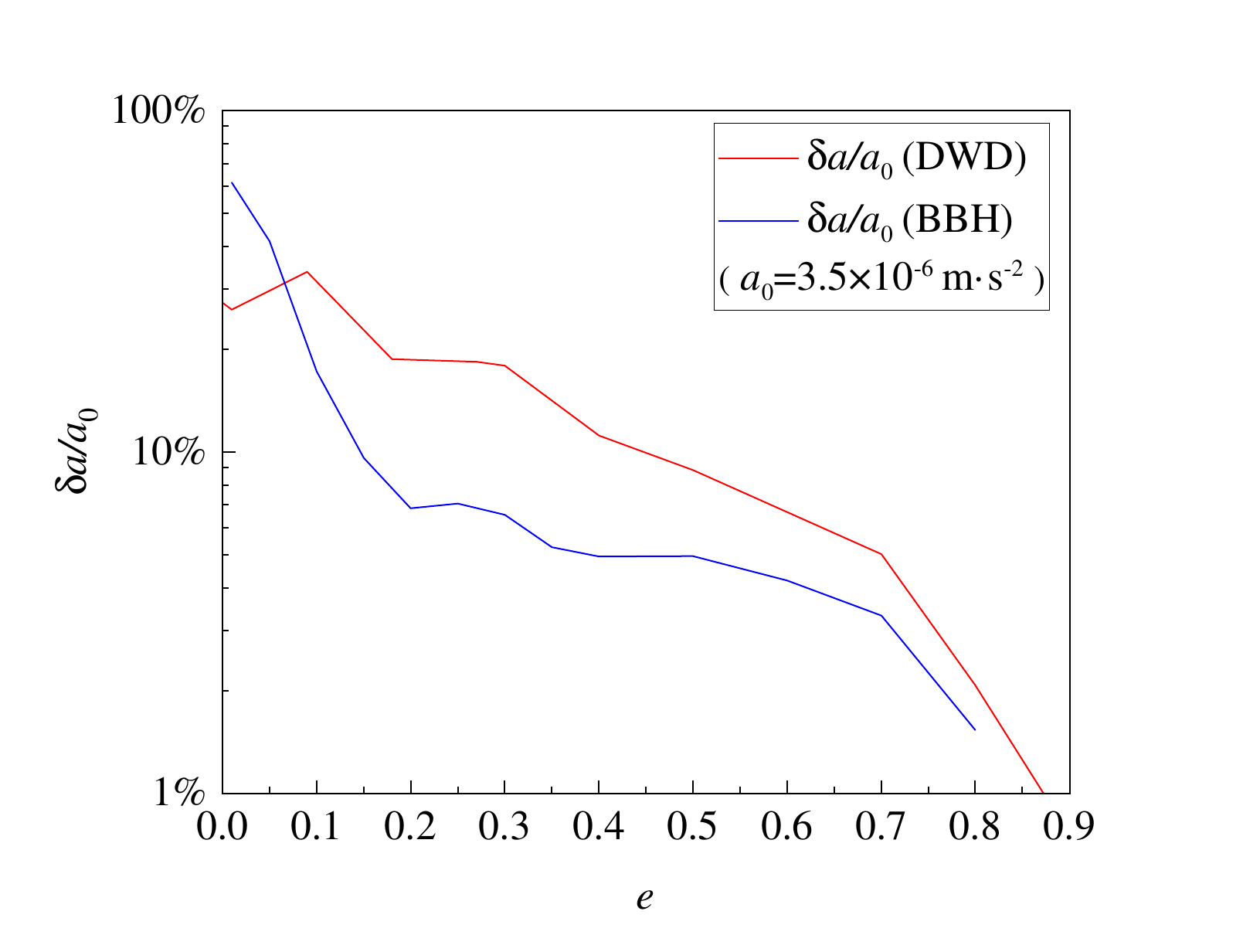}
\caption{{\bf Relative error of acceleration measurement as a function of the compact binary's eccentricity.} Here we place a compact binary system at $0.4$~pc from a $4\times10^{6}\rm M_{\odot}$ SMBH, set the initial GW frequency $f_{e}^{j=2}=3\rm mHz$, observational time $\tau_{{\rm obs}}=4\rm yrs$ and SNR$=20$. We plot the relative error of acceleration measurement, as a result of the fisher matrix analysis (see text), as a function of binary's initial eccentricity. Red line shows a DWD system with $m_{1}=0.35 {\rm{M}}_{\odot}, m_{2}=0.4 {\rm{M}}_{\odot}$, while blue line represents a BBH system with $m_{1}=15 {\rm{M}}_{\odot}, m_{2}=20 {\rm{M}}_{\odot}$. Note that the amplitude of peculiar acceleration is $a_{0}\sim 3.5\times 10^{-6} m\cdot s^{-2}$ for this configuration.}
\label{fig:relative_accuracy}
\end{figure}

Next (Figure~\ref{fig:mismatch_bbh}) we consider a BBH system orbiting around a SMBH. As mentioned in Section~\ref{Analytical 2}, since the BBH system has a larger chirp mass than DWD and DNS systems, its intrinsic chirp rate changes significantly during the observation (See Eq.~(\ref{eq:chirp mass})). The varying chirp rate makes it hard to estimate analytically if the acceleration is distinguishable. However, the numerical analysis (see Section~\ref{NumericalB}) is still valid and can be used to estimate if we can distinguish an accelerating signal from a non-accelerating signal in this case. 

In Figure \ref{fig:mismatch_bbh}, we fix the BBHs' distance to the SMBH as $0.1\rm pc$ and vary the eccentricity of the BBH system. As shown in this Figure, the mismatch between the accelerating GW signal and its optimal fit in non-accelerating templates grows with eccentricity. In particular, the signal-to-noise ratio requirement for distinguishing the acceleration drops from $130$ to $10$ as the eccentricity increases from $0$ to $0.7$. This Figure is also a proof of concept that distinguishing an accelerating GW source from a non-accelerating one can be done for an eccentric system with a large chirp mass, i.e., BBH.  

\subsection{Measuring the Acceleration -  Numerical Approach}
\label{egB}

With the help of Fisher Matrix Analysis (See Section~\ref{NumericalC}), we can give a numerical estimation of the acceleration measurement accuracy, providing that GW templates with acceleration are adopted in the parameter extraction. 
\begin{figure}[htbp]
\centering
\includegraphics[width=3.5in]{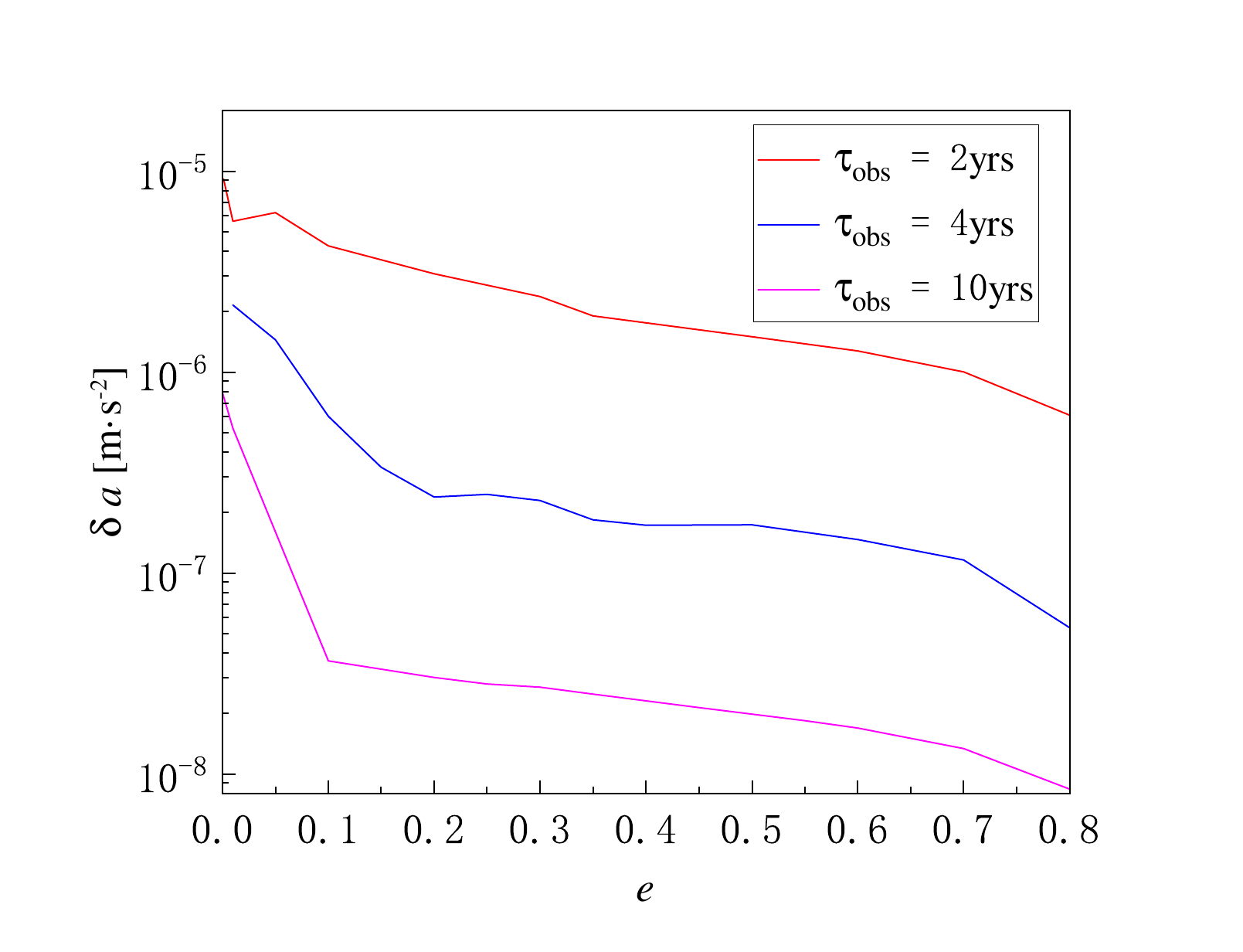}
\caption{{\bf Absolute error of acceleration measurement as a function of BBHs' eccentricity for different observation durations.} Here we take the same BBH as in Figure \ref{fig:relative_accuracy}, but compare the absolute error of acceleration measurement for different observation durations (2yrs/ 4yrs /10yrs). The initial frequency of GW is $f_{e}^{j=2}=3\rm mHz$ and SNR is set to be 20 for all the cases.}
\label{fig:accuracy_bbh2}
\end{figure}
Figure~\ref{fig:relative_accuracy} demonstrates how the relative error of acceleration measurement ($\delta a / a_0$, where $a_0$ is the intrinsic acceleration) decreases with increasing eccentricity. As a proof of concept, we fix the signal-to-noise ratio ($\rm{SNR}=20$), observation duration ($\tau_{\rm obs}=4yrs$), and the acceleration ($a=3.5\times 10^{-6}m \cdot s^{-2}$). As shown in the Figure, the error of acceleration measurement is quite large ($\sim 100\%$) when the orbit eccentricity is zero. However, the error sharply drops to $\sim5\%$  for moderate eccentricity and is lower than $1\%$ when the eccentricity is over $0.8$. This sharp decrease happens because the multiple harmonics in the GW signal can be detected even when the binary's eccentricity is $\sim0.1$. This feature enables us to detect the eccentric GR precession pattern and disentangle the acceleration from the chirp mass measurement.

Additionally, the acceleration measurement accuracy's dependence on eccentricity varies for different kinds of compact binaries. In particular, BBHs have a larger error of acceleration measurement than DWDs when their eccentricity is negligible. However, the error drops more sharply along with the increase of BBHs' eccentricity. This fact is consistent with the analysis in Section~\ref{sec:MeasureAcc}: the large chirp mass of BBHs renders its large intrinsic chirp rate $\dot{f}_{e}$. Therefore, because of the degeneracy between the acceleration and the chirp mass ($\dot{f}_{acc}$ and $\dot{f}_{e}$), BBHs will have larger uncertainty of acceleration if the eccentric precession pattern is not detected and $\dot{f}_{e}$ is not disentangled. However, when the eccentricity increases and the precession pattern helps us disentangle the acceleration, the large chirp rate of BBH enables us to get a higher measurement accuracy of the source's parameters, including acceleration. 
\begin{figure*}[t!]
\centering
\includegraphics[width=0.47\linewidth]{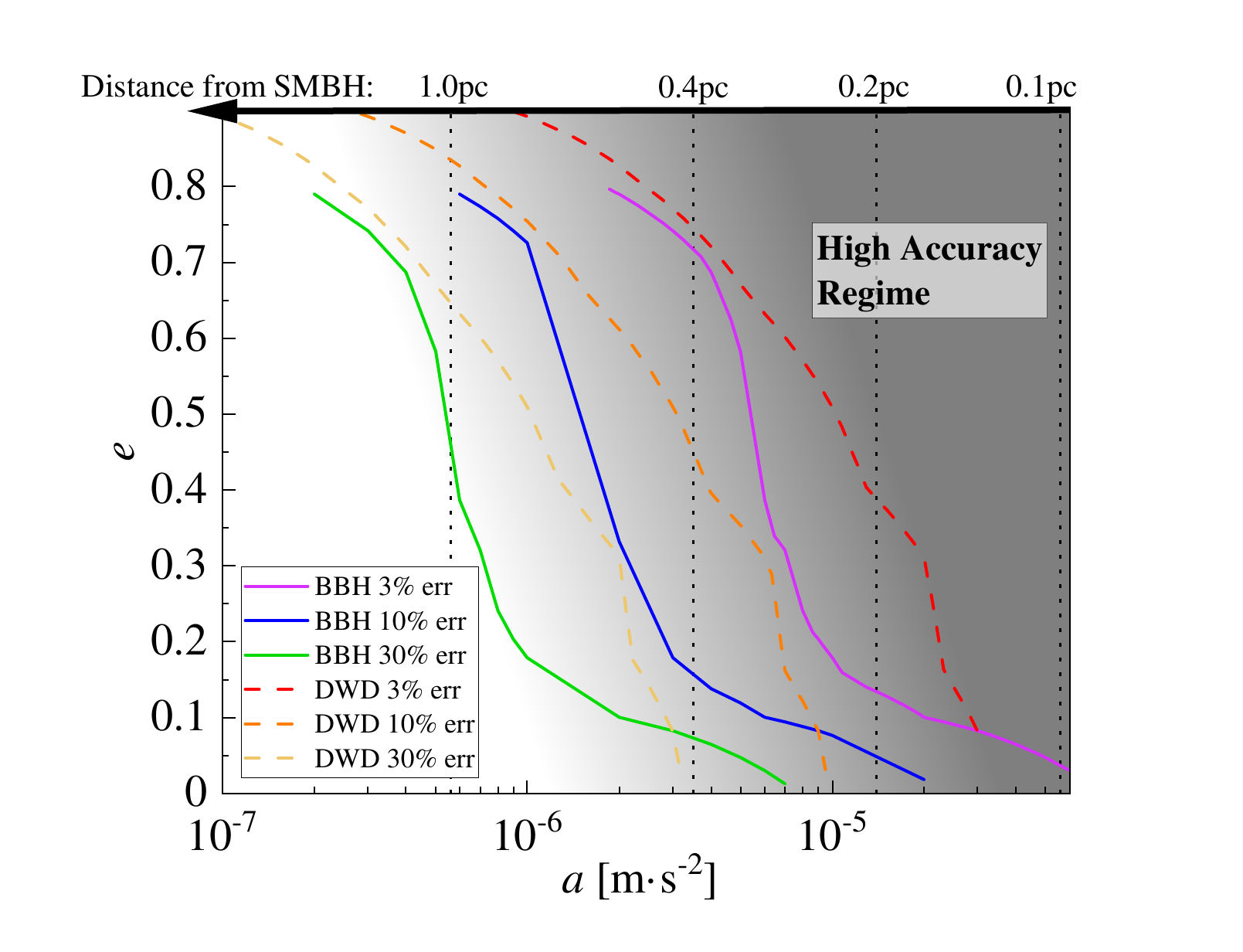}
\includegraphics[width=0.47\linewidth]{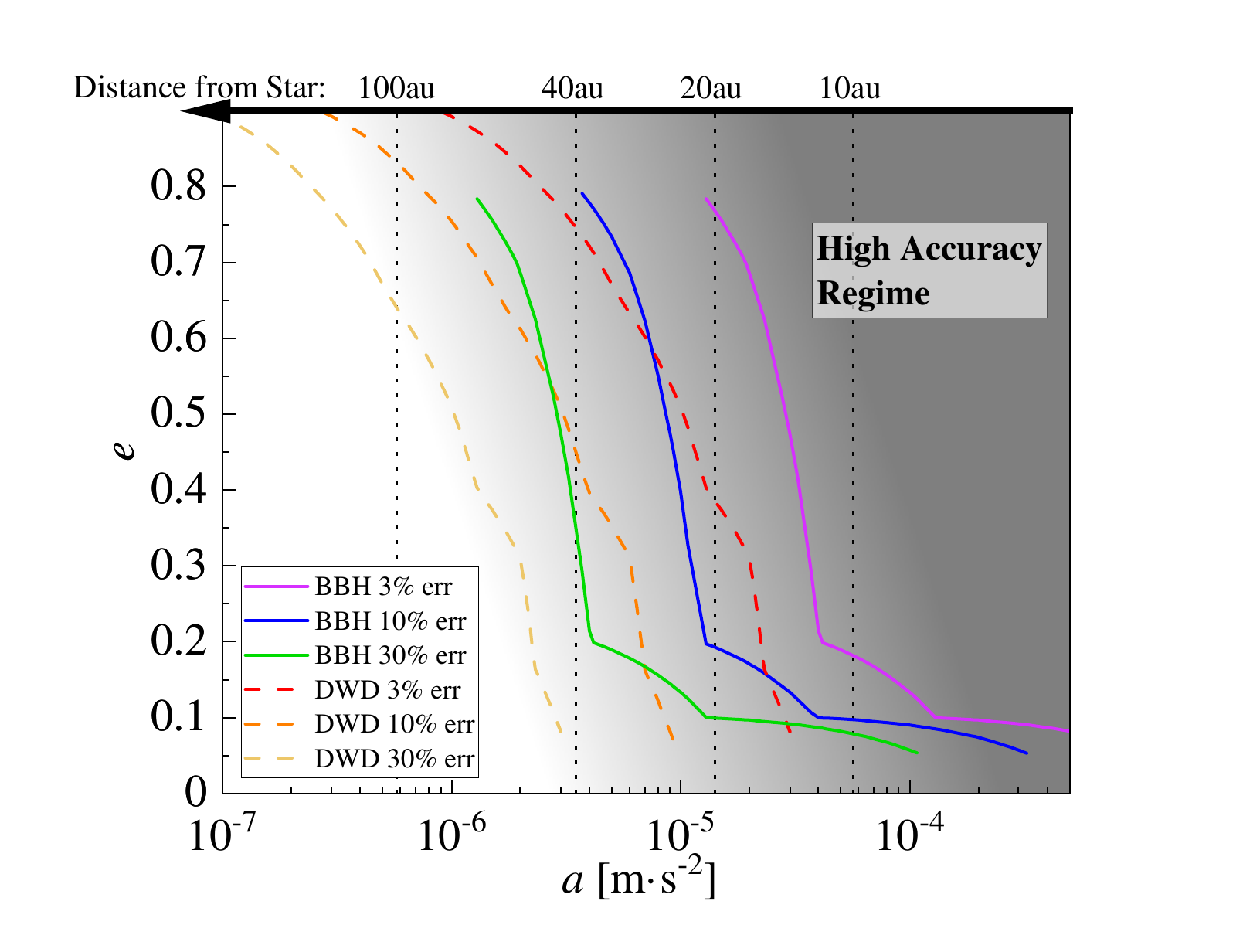}
\caption{{\bf Critical eccentricity, for the measurement accuracy of acceleration to reach given thresholds, as a function of the binary's peculiar acceleration.} For each line in the two panels, we set a targeted accuracy of peculiar acceleration measurement (i.e., $30\%,10\%$, and $3\%$, from left to right), and show the critical eccentricity of the binary for GW measurement to reach such accuracy as a function of the binary's acceleration. The solid lines represent a BBH ($m_{1}=15 {\rm{M}}_{\odot}, m_{2}=20 {\rm{M}}_{\odot}$), and the dashed lines represent a DWD ($m_{1}=0.35 {\rm{M}}_{\odot}, m_{2}=0.4 {\rm{M}}_{\odot}$). Both systems have the initial GW frequency $f_{e}^{j=2}=3\rm mHz$. 
As explained in the text, the acceleration measurement can be applied to a wide range of tertiaries. Here, as an example we consider a $4\times 10^{6}\rm M_{\odot}$ SMBH case ({\it Left Panel}), and a $1 \rm M_{\odot}$ star case ({\it Right Panel}). We set the observation duration as $4$~yrs and SNR to be $20$, compute the absolute error of acceleration measurement for a fixed distance from the tertiary ($r=0.4pc$ for the SMBH case, $r=100au$ for the star case), and analytically generalize the results to get the relative error for the different magnitude of $a$ (distance to the tertiary). The acceleration measurement exceeds the targeted accuracy in the region to the right of each line, where we darken the background color to highlight the difference. For illustration purposes, we translate the acceleration into the binary's distance to the tertiary and mark it on the top of each panel. This figure highlights that higher eccentricity can extend the parameter space where acceleration can be accurately measured.
}
\label{fig:ecc_acc}
\end{figure*}

Figure~\ref{fig:accuracy_bbh2} demonstrates the effect of the observation duration on the acceleration measurement accuracy. Here we focus on the case of an eccentric BBH system around a SMBH and adopt the same configuration as in Figure~\ref{fig:relative_accuracy} to quantify LISA's ability of acceleration measurement. The signal-to-noise ratio is set to be $20$, and we consider three characteristic observation duration ($\tau_{\rm obs}=2,4$ and $10$ years, from top to bottom, respectively). For a stellar mass BBH system like the one in Figure~\ref{fig:accuracy_bbh2}, the accuracy of acceleration measurement is around $\sim10^{-7}m\cdot s^{-2}$ for a 4-year mission, and can be $\sim10^{-8}m\cdot s^{-2}$ for a 10-year mission.

We emphasize that the period of the outer orbit ($\sim10^{5}yrs$) is much longer than the observation duration ($\sim 4yrs$) in Figure \ref{fig:accuracy_bbh2}. Thus, the results of this figure can be generalized to other cases when the acceleration is almost constant during the observation. 

Figure~\ref{fig:ecc_acc} maps the parameter space where we can accurately measure the binary's peculiar acceleration. In particular, the left panel corresponds to a compact binary orbiting around the $4\times10^{6}M_{\rm \odot}$ SMBH, and the right panel demonstrates the case when the tertiary is a $1M_{\rm \odot}$ star. Setting the x-axis as the amplitude of acceleration (distance to the tertiary) and the y-axis as the eccentricity of the inner binary, we plot the boundary of the regimes where the acceleration measurement accuracy is higher than the given values. The result shows that LISA can measure an eccentric compact binary's peculiar acceleration when it is within $\sim10^{0}\rm pc$ from a $4\times 10^{6}\rm{M}_{\odot}$  SMBH, or $\sim 10^{2}\rm AU$ from a stellar-mass tertiary.

As shown in Figure~\ref{fig:ecc_acc}, the existence of eccentricity extends the parameter space where the binary's acceleration can be measured, allowing us to identify more distant tertiaries (up to $\sim10$ times the distance when the inner binary is circular). Moreover, when the binary's eccentricity is small, DWDs have a higher accuracy of acceleration measurement than BBHs. However, the increase in eccentricity can enhance the BBHs' acceleration measurement more significantly, which is consistent with our analysis of Figure \ref{fig:relative_accuracy}.

\section{Discussion}
\label{conclusion}
Many LISA-band compact object binaries have non-negligible acceleration caused by the gravitational pull of either a nearby stellar-mass tertiary or a SMBH in the galactic center \citep[e.g.,][]{Antognini+14,antognini16,Hoang+18,Stephan+16,Stephan+19,Wang+21,Rose+20}. Moreover, these accelerating GW sources are likely to have eccentricity in the millihertz band 
\citep{O'Leary+09,Samsing+18,Hoang+19,Fragione+19,breivik20,Naoz+20,Wang+21,Zhang+21}. In particular, in some dynamical channels, the existence of a tertiary can directly produce LISA-band sources by exciting the eccentricity of inner orbit and accelerating the merger \citep{Thompson+11,Antognini+14,Hoang+18,Stephan+19,Martinez+20,Hoang+20,Naoz+20,Stephan+19,Wang+21}. Thus, we expect the accelerating eccentric binaries to be a common feature in the future GW observation. 

However, it may be difficult to ascertain the properties of these sources because a degeneracy exists between the chirp-mass-induced frequency shift and
acceleration-induced frequency shift \citep{Robson+18,Tamanini+20,chen20envi,Xuan+21}. This degeneracy happens because both the chirp mass and the peculiar acceleration contribute to the frequency shift rate in the GW signal for the leading order, making it hard to disentangle chirp mass from acceleration in data analysis. 

In this work, we explored the detection of peculiar acceleration for eccentric compact binaries in the LISA band, taking into account the effect of GR precession pattern. We find that the eccentric GR precession pattern can break the degeneracy between the compact binary's acceleration and chirp mass (See the flowchart in Figure~\ref{fig:flowchart} for our overall approach).  Therefore, it can be much easier to detect the acceleration of eccentric GW sources than circular ones.

By deriving analytical formulas, we quantified how eccentricity helps with distinguishing the accelerating GW sources from non-accelerating GW templates (see Eq.(\ref{eq: criteria})).
Furthermore, we demonstrated that the existence of eccentricity can improve the measurement accuracy of acceleration. Analytically, we constrained such accuracy for DWDs in Figure \ref{fig:dwd-ana}.

The analytical results are verified numerically, and the ability of LISA to detect the peculiar acceleration of eccentric GW sources is estimated. For example, we find that the critical distance to distinguish the acceleration is $\sim2\rm pc$ for the case of a DWD system orbiting around a $4\times 10^{6}\rm{M}_{\odot}$ SMBH, with $e=0.1$ and $\rm{SNR}\sim20$ (i.e., Figure~\ref{fig:mismatch_dwd}). 
Additionally, we find that the critical $\rm SNR$ for distinguishing the acceleration of a stellar-mass BBH system can drop from $130$ to $10$ when the eccentricity rises from $0$ to $0.7$ (for example, as highlighted in Figure~\ref{fig:mismatch_bbh}).
Moreover,  by adopting the GW templates with acceleration, we can improve the accuracy of acceleration measurement by a factor of $10\sim100$ for GW sources with moderate eccentricity, compared with the zero-eccentricity cases (see Figure~\ref{fig:relative_accuracy}).

During a 4-year LISA mission, the accuracy of acceleration measurement can be $\sim 10^{-7} m\cdot s^{-2}$ when $\rm SNR=20$ for an eccentric stellar-mass BBH system orbiting around a SMBH. Such accuracy enables us to measure the binary's acceleration when it is $\sim1\rm pc$ from a $4\times 10^{6}\rm M_{\odot}$ SMBH. Moreover, the accuracy of acceleration measurement for eccentric compact object binaries can be even higher when the observation duration is longer (see Figure~\ref{fig:accuracy_bbh2}), or if the eccentricity of the GW source is higher (as depicted in Figure~\ref{fig:ecc_acc}).

Our results highlight the importance of eccentricity in GW astronomy, as it can disentangle the compact binary's parameters and significantly enhance acceleration measurement. In future GW data analysis, adopting GW templates that include both eccentricity and acceleration can be meaningful to our understanding of the environment of the GW sources.
\\
\acknowledgments
ZY acknowledges the partial support from the Mani L. Bhaumik Institute for Theoretical Physics summer fellowship and a partial support of the UCLA's  Summer Mentored Research Fellowship (SMRF). SN acknowledges the partial support from NASA ATP 80NSSC20K0505,  NSF through grant No.~2206428,  and thanks Howard and Astrid Preston for their generous support. XC is supported by the Chinese National Science Foundation through grant No.~11873022.
\appendix

\bibliographystyle{apsrev4-1.bst}
\bibliography{mybib,bibbase}

\end{document}